\documentclass{aastex63}

\usepackage{booktabs}

\begin{document}
\title{Mass--accretion, spectral, and photometric properties of T Tauri stars in Taurus based on TESS and LAMOST}

\author[0000-0001-5989-7594]{Chia--Lung Lin}
\affiliation{Graduate Institute of Astronomy, National Central University, Taoyuan 32001, Taiwan}

\author[0000-0002-3140-5014]{Wing--Huen Ip}
\affiliation{Graduate Institute of Astronomy, National Central University, Taoyuan 32001, Taiwan}
\affiliation{Graduate Institute of Space Science, National Central University, Taoyuan 32001, Taiwan}

\author{Yao Hsiao}
\affiliation{Graduate Institute of Astronomy, National Central University, Taoyuan 32001, Taiwan}

\author{Tzu-Hueng Chang}
\affiliation{Graduate Institute of Astronomy, National Central University, Taoyuan 32001, Taiwan}

\author[0000-0001-7255-5003]{Yi--han Song}
\affiliation{Key Laboratory of Optical Astronomy, National Astronomical Observatories, Chinese Academy of Sciences, Beijing 100012, China}

\author[0000-0001-7865-2648]{A--Li Luo}
\affiliation{Key Laboratory of Optical Astronomy, National Astronomical Observatories, Chinese Academy of Sciences, Beijing 100012, China}

\received{2022 October 20}
\revised{2023 May 8 (R1) \& 2023 June 16 (R2)}
\accepted{2023 June 28}
\submitjournal{Astronomical Journal}




\begin{abstract}
We present the analysis of 16 classical~T~Taur~stars using LAMOST and TESS data, investigating spectral properties, photometric variations, and mass-accretion rates.
All 16 stars exhibit emissions in H$\alpha$ lines, from which the average mass-accretion rate of $1.76\times10^{-9}~M_{\odot}yr^{-1}$ is derived.
Two of the stars, DL~Tau and Haro~6--13, show mass--accretion bursts simultaneously in TESS, ASAS--SN, and/or ZTF survey.
Based on these observations, we find that the mass--accretion rates of DL~Tau and Haro~6--13 reach their maximums of $2.5 \times 10^{-8}~M_{\odot}yr^{-1}$ and $2 \times 10^{-10}~M_{\odot}yr^{-1}$ during the TESS observation, respectively.
We detect thirteen flares among these stars.
The flare frequency distribution shows that the CTTSs' flare activity is not only dominated by strong flares with high energy but much more active than those of solar-type and young low--mass stars.
\textnormal{By comparing the variability classes reported in the literature, we find that the transition timescale between different classes of variability in CTTSs, such as from Stochastic (S) to Bursting (B) or from quasi-periodic symmetric (QPS) to quasi-periodic dipping (QPD), may range from 1.6 to 4 years.}
We observe no significant correlation between inclination and mass--accretion rates derived from the emission indicators.
This suggests that inner disk properties may be more important than that of outer disk.
\textnormal{Finally, we find a relatively significant positive correlation between the asymmetric metric \emph{M} and the cold disk inclination compared to the literature.}
A weak negative correlation between the periodicity metric \emph{Q} value and inclination has been also found.

\end{abstract}
\keywords{stars: evolution -- stars: young stellar objects -- stars: low-mass stars -- open clusters and associations: Taurus Molecular Cloud}


\section{Introduction}



%
%
Classical T Tauri stars (CTTSs) are young, low-mass pre--main--sequence objects with typical spectral types between F and M, that are accreting material from the circumstellar disk.
These stars exhibit strong emission lines in their spectra and show an excess of emission ranging from radio to ultraviolet relative to the stellar photosphere.
The infrared excess emission from CTTS indicates the presence of a circumstellar disk \citep[][]{1968ApJ...151..977M, 1974MNRAS.168..603L, 1987ApJ...312..788A, 1988ApJ...330..350B} and can be used to infer the amount of dust and the evolutionary status of the system.
The amount of circumstellar material drops as the protostellar system evolves, leading to the decrease of the excess emission in the infrared.

The CTTSs have luminosity variations on different timescales, from hours to even decades \citep[][]{2008MNRAS.391.1913R, 2013AJ....145...79C}, depending on various physical phenomena affecting the stellar and circumstellar environment.
\textnormal{For example, the obscuration of the central star’s photosphere by circumstellar dust may lead to an aperiodic dip of light \citep[e.g.,][]{2015A&A...577A..11M}.
The presence of stable starspots on the stellar surface may result in a periodic modulation in the light curve of the star \citep[e.g.,][]{2018ApJ...862...44K}.
The unsteady accretion processes and magnetic activity (i.e., flares) may cause the random and short timescale brightening variability.
The stochastic light curve of the CTTS may be attributed to the superposition of the various types of variability resulting from different mechanisms mentioned earlier. All these mechanisms can vary with time.
}

One of the major sources of variability in CTTSs during this evolutionary phase is the accretion--driven burst \citep[e.g.,][]{2022arXiv220311257F}.
\textnormal{
There are various types of accretion--driven bursts that last for different timescales.
One type is the FU~Ori--like outburst, which is characterized by a large increase in luminosity (typically by a factor of tens) over a timescale of decades to hundreds of years \citep[e.g.,][]{2005MNRAS.361..942C}.
In addition, the outbursts with the type of V1647~Ori, EX~Lup, and Protostellar share the similar timescale of a few years with a luminosity increase ranging from 1 to 5 mag.
Lastly, the one we focus on in this study is the accretion bursts with a luminosity variability up to a few magnitudes on a relatively short timescales from a few hours to days at optical wavelengths \citep[e.g.,][]{2018AJ....156...71C}.
}
The numerical simulations carried out by \citep[e.g.,][]{2020MNRAS.491.1057C} suggest that this variability could be caused by unstable accretion processes.
In addition, such events, coupled with rotational modulation, tend to dominate changes in the accretion rate \citep[][]{2017MNRAS.465.3889R,2020MNRAS.491.5035S}.

According to the magnetospheric accretion mechanism \citep[e.g.,][]{1990RvMA....3..234C}, the most viable scenario currently used to describe the accretion behaviors of CTTSs, the central star is channeling the material in the circumstellar disk to the stellar surface through the magnetospheric accretion columns \citep[][]{1999ARA&A..37..363F, 2003ApJ...592..266M}.
\textnormal{The gas is heated to a temperature of $10^{4}$--$10^{6}$~K when the channeled materials hit the stellar surface, creating a shock on the stellar photosphere, which emits intense X--rays \citep[][]{2007A&A...465L...5A, 2007A&A...466.1111G,2017A&A...597A...1C}.}
The X--rays are mostly reabsorbed by the accretion columns, from which the lower energy photons, mainly in blue band, are re--emitted, resulting in the luminosity excess from visible to ultraviolet wavelength.
\textnormal{Thus, the U--band excess has been often used to determine the accretion luminosity, from which the mass--accretion rate can be determined \citep[][]{1998ApJ...492..323G,2008ApJ...681..594H}.}
In addition, such an accretion mechanism also produces detectable emission features in CTTSs spectra, such as H$\alpha$ and H$\beta$ \citep[e.g.,][]{1989ApJ...341..340B}.
\textnormal{\cite{2016A&A...586A..47S} shows that when U--band excesses are not available, mass-accretion rates can be obtained with good results from the H$\alpha$ flux using the currently available calibrations from the literature.}
Therefore, these emission features can be used to probe the \textnormal{state (i.e., active or inactive)} of the mass--accretion process.
\textnormal{
Furthermore, multi--band simultaneous observations can be useful for estimating the mass-accretion rate of CTTSs even without direct U--band measurements.
For example, \cite{2017ApJ...835....8T} and \cite{2018ApJ...862...44K} used optical wavelength observations of accretion burst variability to estimate the mass-accretion rate of DQ Tau.
\textnormal{In \cite{2018ApJ...862...44K}}, they used the Kepler light curve to identify the burst phenomenons and extrapolated the associated U--band excess luminosities based on the simultaneous BVRI observations of the star.
}

In this study, we present mass--accretion rates and the properties that can be determined with our data for \textnormal{selected} CTTSs in the Taurus Molecular Cloud (hereafter, TMC), which is a nearby star--forming region located at a distance about 147~pc from the Sun \textnormal{\citep[][]{2020A&A...633A..51Z}} with an age of 1--3~Myr \citep[][]{2017ApJ...844..158S}, using high photometric precision light curves and low--resolution spectra from the space--based telescope, TESS (the Transiting Exoplanet Survey Satellite) and LAMOST (the Large Sky Area Multi--Object Fibre Spectroscopic Telescope), respectively.
\textnormal{Given the lengths of the available light curve data, we will primarily focus on the short timescale burst.}

The outline of this paper is as follows. 
In Section~\ref{sec:section2}, we introduce the TESS and LAMOST observations of TMC members, including \textnormal{their stellar} parameters reported in literature.
In Section~\ref{sec:section3}, we explain our methodology for estimating the mass--accretion rates of our CTTS targets \textnormal{using various emission indicators from LAMOST low-resolution spectra and TESS light curves.}
We also detail our method for determining photometric variabilities in this section.
In Section~\ref{sec:section4}, we describe our results of the data analysis and implications.
Finally, in Section~\ref{sec:section5}, we summarize the result and conclude the study.

\section{Sample and data sets}
\label{sec:section2}

\subsection{LAMOST spectroscopic survey on T Tauri stars}
LAMOST, the Large Sky Area Multi--Object Fibre Spectroscopic Telescope located at the Xinglong Observatory of the National Astronomical Observatories of Chinese Academy of Sciences (NAOC), is a reflecting Schmidt telescope with an effective aperture of $\sim4$~m and a wide field of view of 20 deg$^{2}$ in the sky \citep[][]{2012RAA....12..723Z}.

With 4000 fibers that are mounted on the focal plane, it has been conducting spectroscopic surveys efficiently for stars, galaxies and quasars mostly in the northern celestial hemisphere above the declination of $-10$~degree since Autumn 2012.
\textnormal{Each fiber has a scale of 3.3~arcseconds on the focal surface. 
The centers of the spectroscopic fibers are about 4.47 arcminutes apart, and each fiber can move at most 3~arcminutes from its center position.
Sixteen spectrographs are used in the system, and each receives the dispersed light from 250 fibers.
The resolution of LAMOST spectra is R~$\approx$~1800 around the sdss--g band with a wavelength coverage of 3700--9100~\AA.}

\textnormal{
The LAMOST~2D~pipeline is utilized for data reduction. Initially, each exposure's data from each CCD chip is separately reduced, and then the results from each exposure are combined. Bias and dark are subtracted from each raw image. The flat-field spectrum of each fiber is traced, and a polynomial is fitted to the centroid of each fiber's row position. A Gaussian-like profile is assumed for each fiber's profile in the row direction. Using the same trace function, spectra are extracted from all data on the same night, with minor shifts made to align individual frames. The extracted spectra are divided by the one-dimensional flat-field spectrum. The Hg/Cd and Ne/Ar arc lamp spectra are also extracted using the same trace function, including strong sky emission lines to further improve the wavelength solution. The sky spectra are obtained from 20 sky fibers allocated in each spectrograph. After wavelength calibration, all sky spectra are combined into one super-sky spectrum. The super-sky spectrum is scaled to best fit the sky spectrum in each fiber, and then this scaled super-sky is subtracted from each fiber, and the telluric absorption is removed. The science spectra are flux-calibrated by matching the selected flux standard stars and their templates, which are usually of spectral type A or F. Up to five flux standard stars are observed per spectrograph.
}

\begin{figure}
    \centering
    \epsscale{1}
    \plotone{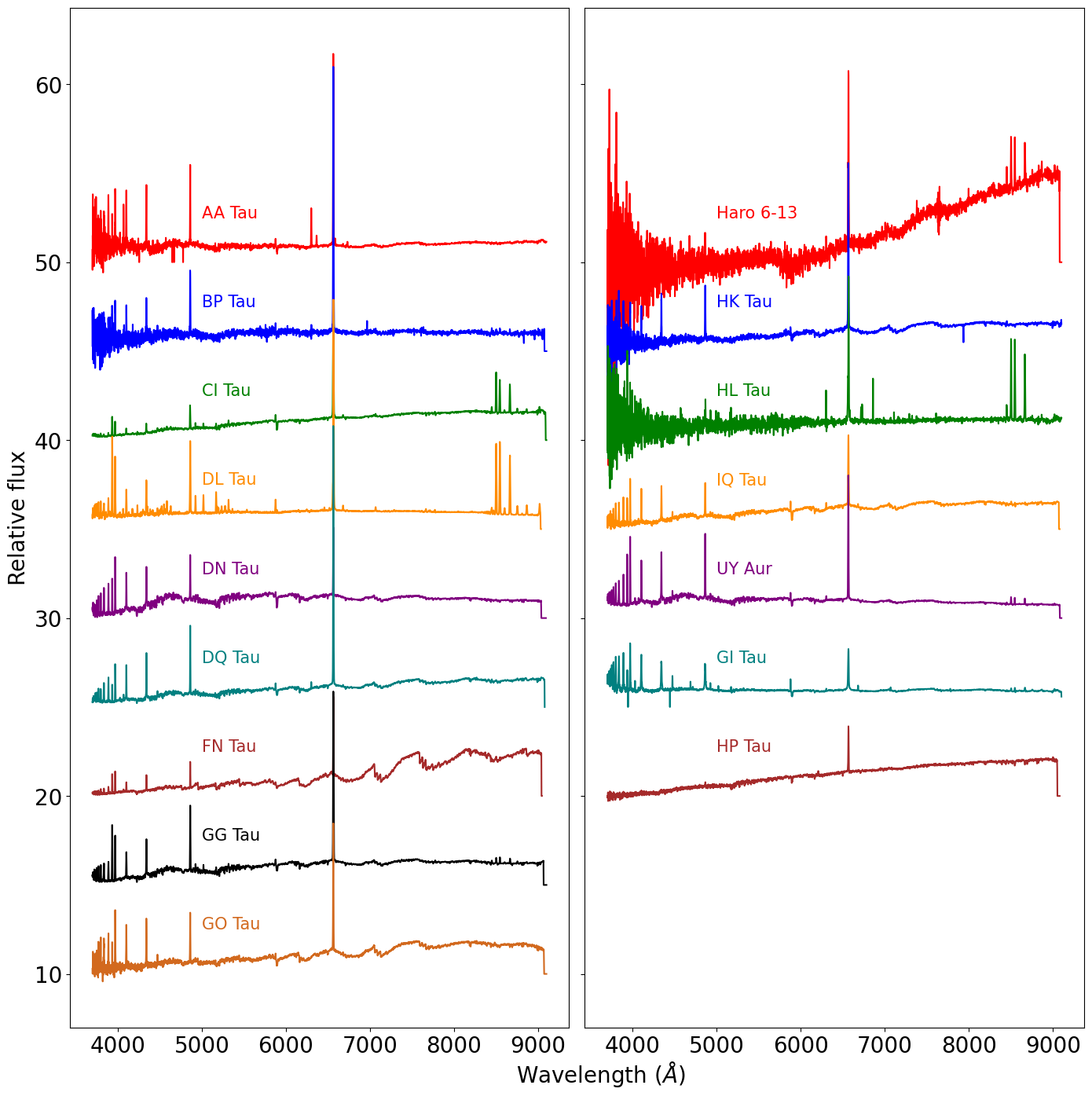}
    \caption{The LAMOST low--resolution spectra of 16 CTTSs in this study}
    \label{fig:figure_lamost_spectra}
\end{figure}

We retrieved spectroscopic data of CTTSs in the TMC from LAMOST's decade--long archive of observational data, and 16 targets are identified (see Table~\ref{tab:tab1}). 
The LAMOST low--resolution spectra of these CTTSs are shown in Figure~\ref{fig:figure_lamost_spectra}.
Table~\ref{tab:tab2} contains the stellar parameters of our targets.
\textnormal{The masses and radii of most CTTSs are provided in \cite{2014ApJ...786...97H}, with the exception of HL~Tau.
The radius (R=7~R$_{\odot}$) and mass (M=1.7M$_{\odot}$) of HL~Tau are given by \cite{Liu et al.(2017)} and \cite{2016ApJ...816...25P}, respectively.
}
The inclinations of \textnormal{the outer disk} have been reported by the adaptive optics \citep[i.e.,][]{1998ApJ...499..883C}, millimeter radio \citep[i.e.,][]{Guilloteau1999}, \textnormal{high--resolution near--infrared imaging \citep[][]{2008ApJ...673L..67K}}, and high--resolution submillimeter resolved image observations of the TMC region \citep[][]{2015ApJ...808L...3A, 2019ApJ...882...49L, Manara et al.(2019),Villenave2020}.
The distance information for the objects studied in this work is from the parallax measurements reported in \cite{2020yCat.1350....0G}.
\textnormal{HK~Tau is a binary system consisting of two companions, HK~Tau~A and HK~Tau~B, with HK~Tau~A being much brighter than HK~TauB ($G$=14.106 vs $G$=17.962 \citep{2021A&A...649A...1G}). 
The disk inclinations of the two companions are highly misaligned, with HK~Tau~A having an inclination of $i$=56.9$^{o}$ \citep{Manara et al.(2019)} and HK~Tau~B having an inclination of $i$=84$^{o}$ \citep{Villenave2020}. 
Therefore, we list their inclinations together in the Table and treat them as an uncertainty in our result comparison in the later sections.}

\subsection{TESS observations of T Tauri stars in TMC}
TESS (the Transiting Exoplanet Survey Satellite) is an MIT--led NASA mission launched on April 18, 2018, to search for nearby transiting exoplanets using the fast--sampling (2--minute) time--series data with the high photometric precision, which is also well suited for studies of stellar physics.
The Taurus Molecular Cloud was in the field of view of TESS in Sectors 43 and 44 from 2021 September 16 to 2021 November 6.
Fourteen of our LAMOST CTTS targets have released TESS 2--min cadence light curve data (see Table~\ref{tab:tab1}).

There are two kinds of flux information contained in the TESS data: Simple Aperture Photometric flux (SAP) and the Pre-search Data Conditioning Simple Aperture Photometric flux (PDCSAP). 
The SAP flux is produced by summing the count rates of the pixels within a customized aperture given by the Science Processing Operations Center (SPOC) pipeline \citep[][]{2016SPIE.9913E..3EJ}.
The SAP flux is often affected by systematic errors such as long term trend, which have been removed from the PDCSAP flux using Co--trending Basis Vectors.
Therefore, the PDCSAP flux is often cleaner than SAP flux.
\textnormal{In this study, we use the PDCSAP flux data with good quality (flag bits QUALITY$=$0) in our analyses.}
Although the long term trends might be the true signal of the stars, the unsteady accretion flows have relatively short timescales and should not be affected significantly by removing the long term trend signal.

\begin{figure}
    \centering
    \gridline{\fig{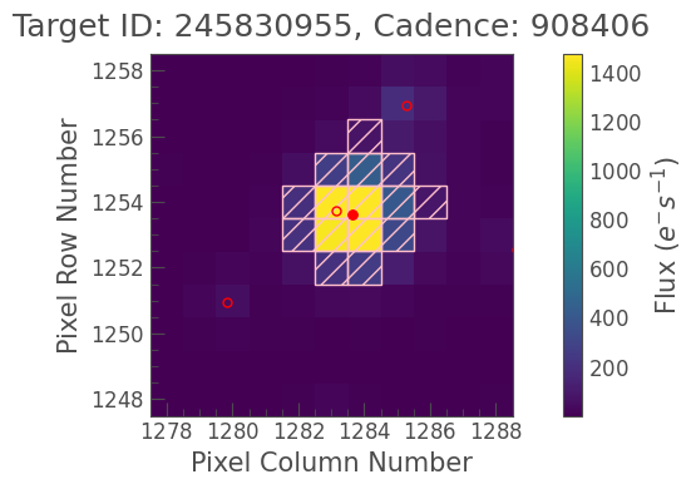}{0.5\textwidth}{} \fig{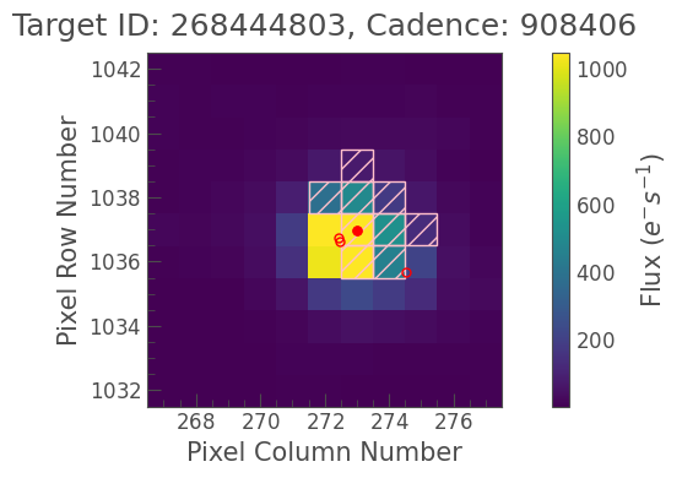}{0.5\textwidth}{}}
    \caption{The example for demonstrating the examination of the contamination from nearby sources. The left panel is the target pixel file (TPF) of GG~Tau (TIC~245830955), and GG~Tau is marked by a filled red circle. The hollow red circles are the field stars with G$_{RP}<$~15 queried from Gaia~DR3. The pink mask pixels are the aperture given by the SPOC pipeline. We found that the field star located in the aperture is much fainter than GG~Tau, and the overall shapes of light curves from different test apertures do not change significantly compared to the SPOC light curve. Therefore, we verified that the data of GG~Tau can be used in this study. The right panel is the TPF of GI~Tau (TIC~268444803). There are three neighbor stars within or at the edge of the SPOC aperture, and one of them is equally bright to GI~Tau, making it hard to distinguish the variability feature between GI~Tau and the neighbor stars in the light curve. As the result, we excluded GI~Tau data from the following analysis.}
    \label{fig:TESS_TPF_example}
\end{figure}

\textnormal{Considering the large size of the TESS pixel scale~($21^{''}\times 21^{''}$), the contamination due to the nearby sources in crowded regions may be unavoidable.
To investigate whether the light curves of our targets of interest are contaminated by the neighboring stars, the Python package Lightkurve \citep[][]{2018ascl.soft12013L} were employed.
We downloaded the target pixel file (TPF) of each star and plotted it together with the coordinates of the field stars with G$_{RP}$~mag~$<$~15 queried from Gaia~DR3 \citep[][]{2020yCat.1350....0G}.
The left panel in Figure~\ref{fig:TESS_TPF_example} is an example, the TPF of GG~Tau (TIC~245830955). 
The pink mask region is the SPOC pipeline aperture, the red hollow circles are the locations of the field stars, and GG~Tau is marked by a red filled circle.
In addition to GG~Tau, there is one neighboring star within the SPOC aperture.
The neighbor star (G$_{RP} \sim 13$) is much fainter than GG~Tau (G$_{RP} \sim 10$).
Also, the overall shapes of the light curves extracted using various test apertures do not change significantly compared to the light curve from SPOC aperture.
As the result, the SPOC given PDCSAP flux of GG~Tau is reliable for the following research.
The right panel in Figure~\ref{fig:TESS_TPF_example} shows another example, GI~Tau (TIC~268444803).
There are three neighbor stars within or at the edge of the SPOC aperture of GI~Tau.
One of them is GK~Tau (G$_{RP}=$~10.89) and is about equally bright to GI~Tau (G$_{RP}=$~11.5).
After the examination with different test apertures, we found that it is impossible to distinguish the variability feature between GI~Tau and the field stars in the light curve.
Therefore, the GI~Tau data was excluded from following analysis.
In the same way, HL~Tau has a similar problem with GI~Tau and was also excluded. 
Ultimately, there were only 12 targets for which the data could be used in the subsequent analyses.}
\section{Analysis}
\label{sec:section3}
\subsection{Calibrating relative flux of LAMOST data to absolute unit}
\textnormal{Due to the limited number of spectral standard stars in the LAMOST survey, no absolute flux-corrected data are available, only relative flux-corrected data can be used.
To obtain the physical flux information, we adopt the template spectra of weak--lines T Tauri stars (WTTS) with the spectral type ranging from K5 to M7 provided by \cite{2013A&A...551A.107M}.
We first normalize the template spectra and LAMOST spectra using the mean flux between 7750 and 8000~$\AA$ because this range has relatively few emission/absorption features.
In this way, we find the best--fit template spectra with the corresponding spectral types for every LAMOST spectra.
We then convert the normalized flux of the template spectra and LAMOST spectra back to the physical flux level of template spectra before normalization.
Finally, the absolute calibrated flux density of LAMOST data (F$_{\lambda, cal}$) is obtained by using the following equation:
\begin{equation}
F_{\lambda, cal} = F_{\lambda, tem}(\frac{d_{2}}{d_{1}})^2,
\label{eq:flux_calibration}
\end{equation}
where F$_{\lambda, tem}$ is the flux density of the best--fit template spectra, the d$_{2}$ is the distance in parsecs of the template star given by \cite{2013A&A...551A.107M}, and d$_{1}$ is the distance in parsecs of our sample.
}

\begin{figure}
    \centering
    \gridline{\fig{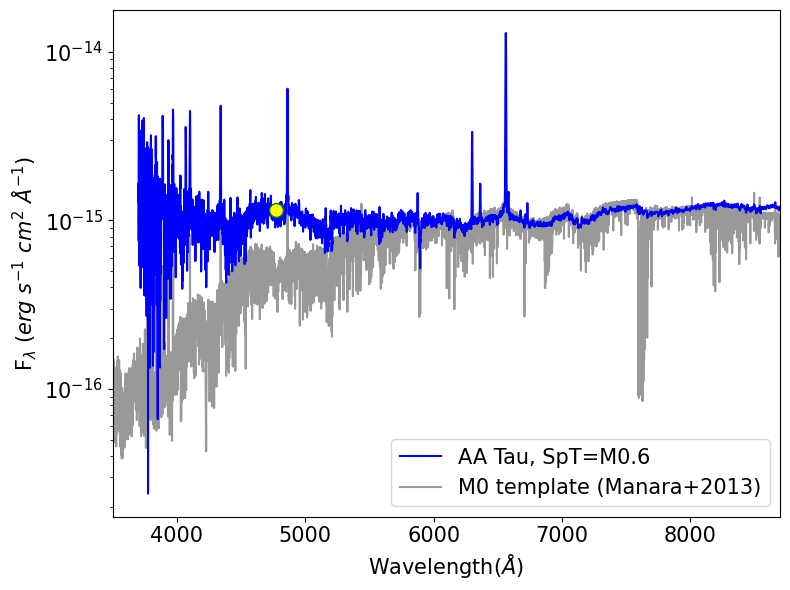}{0.5\textwidth}{} \fig{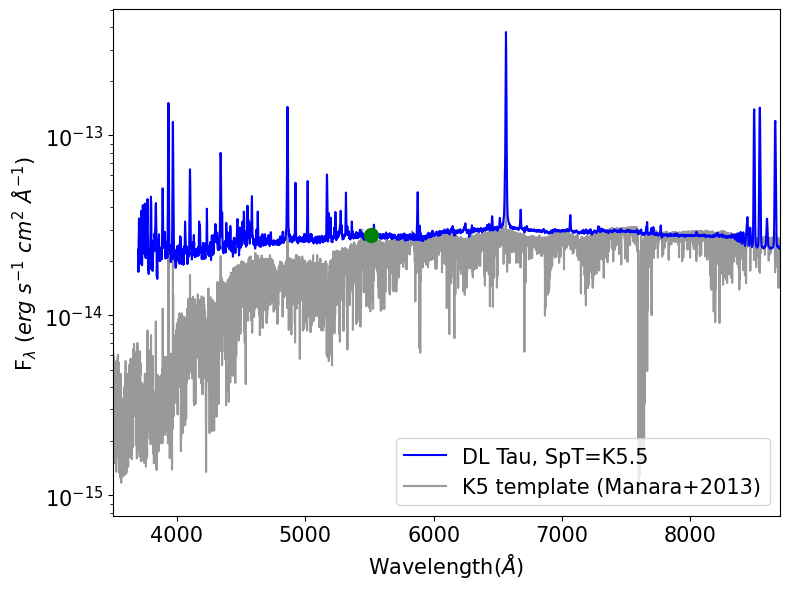}{0.5\textwidth}{}}
    \gridline{\fig{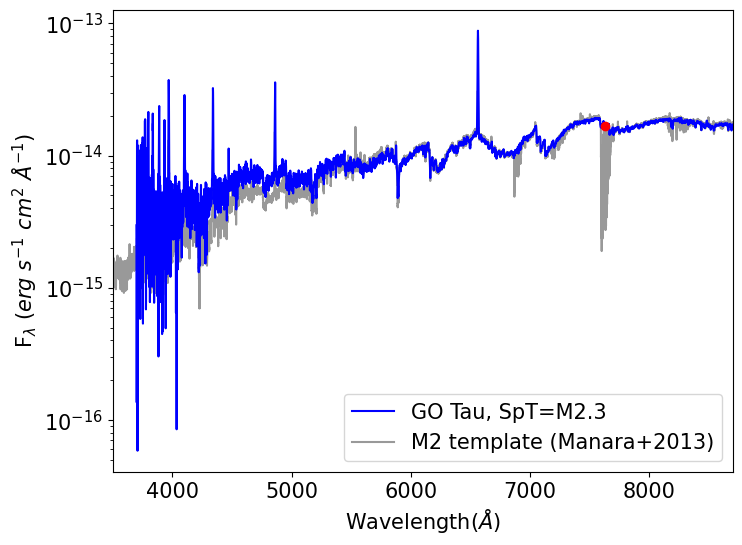}{0.5\textwidth}{}}
    \caption{Three example of the flux calibration processes in this study. The upper left is AA~Tau, the upper right is DL~Tau, and the lower middle is GO~Tau.  In each panel, the blue curve represents the star's spectrum, while the gray curve represents its corresponding best--fit template spectrum given by \cite{2013A&A...551A.107M}. The photometric data from ASAS--SN $g$ band and $V$ band are marked as a yellow circle in AA~Tau case and a green circle in DL~Tau case, respectively. The Pan--STARRS $i$ band data of GO~Tau is marked as a red circle in GO~Tau panel.}
    \label{fig:flux_calibration_spectra}
\end{figure}

\textnormal{To ensure the accuracy of our calibrated absolute flux measurements, we have cross-checked our results with photometric data obtained by ASAS--SN \citep[][]{Jayasinghe et al.(2018), 2019MNRAS.486.1907J} and Pan--STARRS \citep[][]{2016arXiv161205560C, 2020ApJS..251....7F} surveys if any, which were taken within 20 days of the LAMOST observation date. As CTTSs are known to be highly variable stars, this additional step provides further validation of our flux calibration. 
In these surveys, we have identified the following stars in our samples with the photometric data that meet the criteria: (1) from ASAS--SN: AA~Tau in the sdss--g band and BP~Tau in the V~band, (2) from Pan--STARRS sdss--g band: DL~Tau, DN~Tau, Haro~6--13, HL~Tau, and IQ~Tau, and (3) from Pan--STARRS sdss--i band: GO~Tau and HK~Tau.
The flux (in mJy) and observation date (in MJD) of the these photometric data are listed in Table~\ref{tab:tab_asas-sn+panstarr}.
}
\textnormal{
We convert the flux units of the survey photometric data from mJy to the same units as our calibrated LAMOST spectra.
Next, we scale the calibrated LAMOST spectra to match the flux from the surveys at the effective wavelength of the corresponding band. 
This allows us to obtain a more accurate calibrated flux for our LAMOST data.
Figure~\ref{fig:flux_calibration_spectra} shows the example spectra of the flux calibration processes of three stars: DL~Tau, AA~Tau, and GO~Tau.
For stars that lack survey photometric data, we still use their spectra in the following analysis. However, we note that the results obtained from these data could be uncertain by an order of 0 to 1. }

\subsection{Mass--accretion rate measurements from LAMOST spectra analysis}
\begin{figure}
    \centering
    \epsscale{1.1}
    \plotone{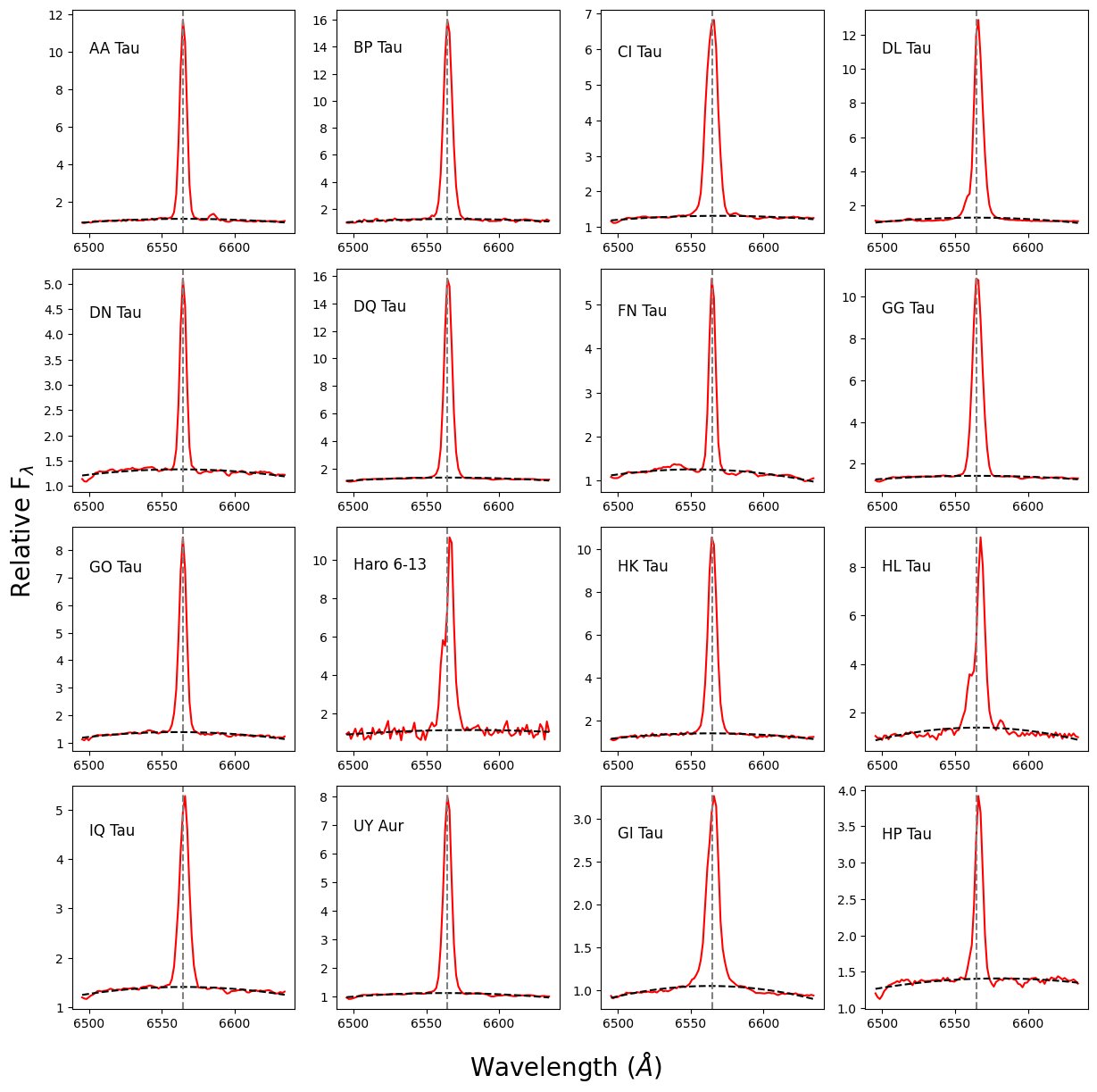}
    \caption{H$\alpha$ emission profiles of our targets observed by LAMOST. In each panel, the spectra is displayed in red, and the black line represents the fit local continuum. The dash vertical line in gray marks the center wavelength of the H$\alpha$ line.}
    \label{fig:figure_emission_profile-ha}
\end{figure}

\begin{figure}
    \centering
    \epsscale{1.1}
    \plotone{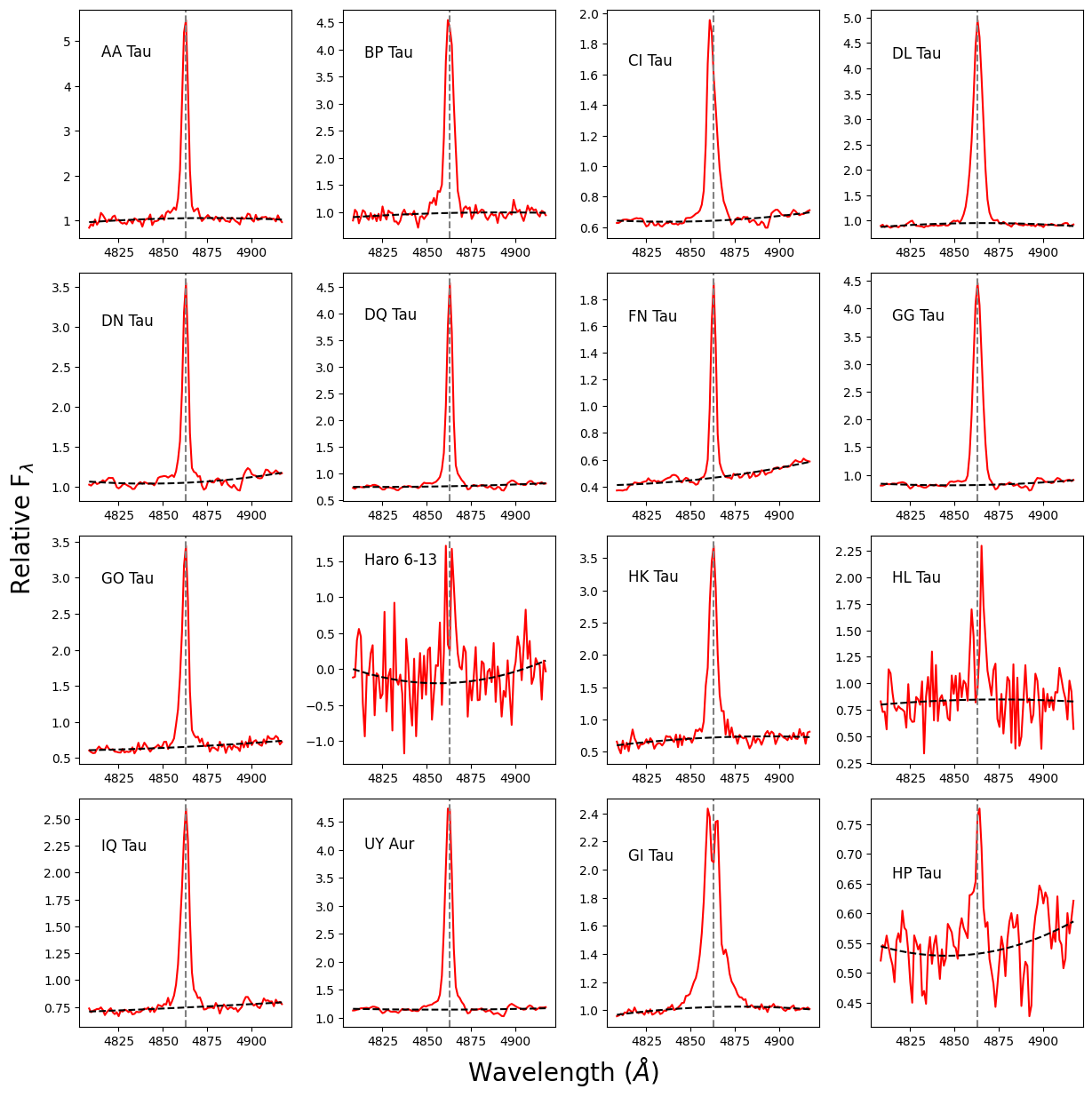}
    \caption{H$\beta$ emission profiles of our targets observed by LAMOST.
    Dashed vertical gray line represents the center of the line in each panel.}
    \label{fig:figure_emission_profile-hb}
\end{figure}

\begin{figure}
    \centering
    \epsscale{0.9}
    \plotone{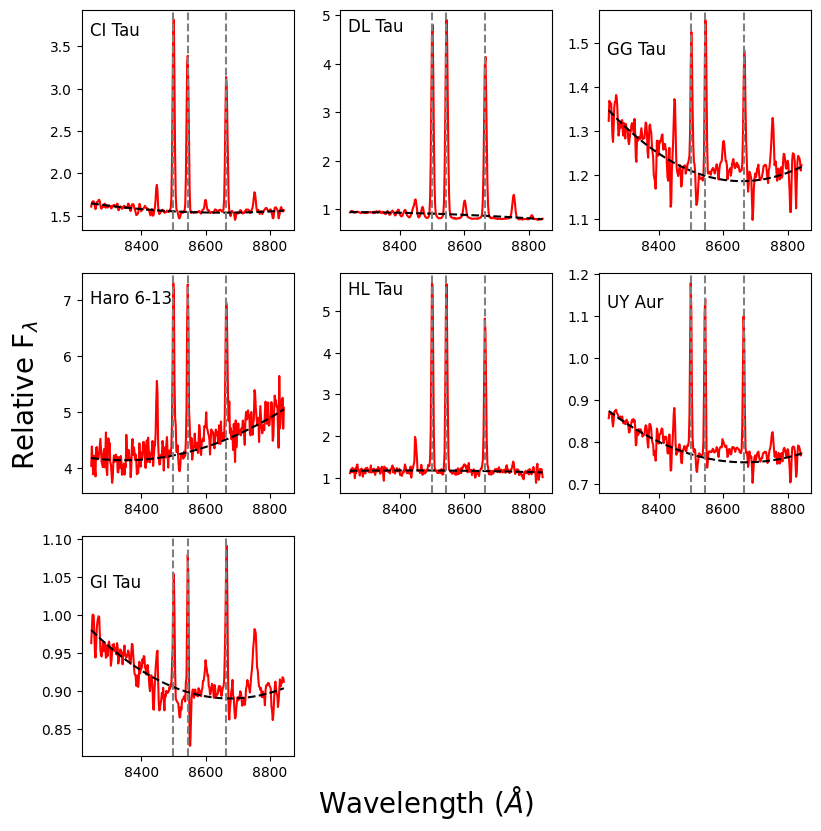}
    \caption{Ca$\text{II}$ 8498, 8542, 8662~$\AA$ emission profiles of our targets observed by LAMOST. Dashed vertical gray lines represent the center of the Ca$\text{II}$ triplet lines in each panel.}
    \label{fig:figure_emission_profile-Ca}
\end{figure}

\begin{figure}
    \centering
    \epsscale{0.9}
    \plotone{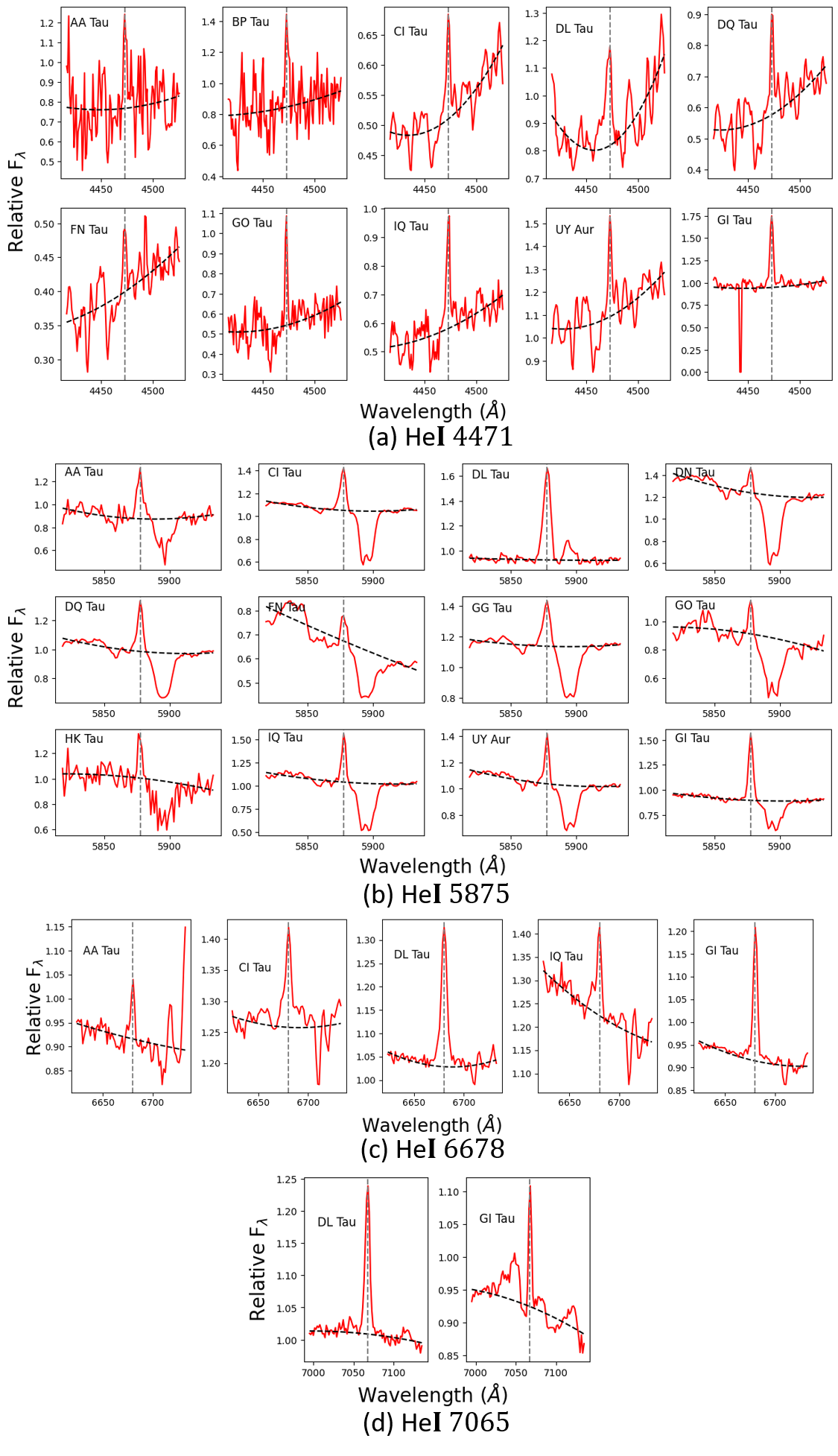}
    \caption{He$\text{I}$ 4471 (a), 5875 (b), 6678 (c), 7065 (d) emission profiles of our targets observed by LAMOST. Dashed vertical gray line represent the center of the He$\text{I}$ in each panel.}
    \label{fig:figure_emission_profile-HeI}
\end{figure}

\textnormal{Using the LAMOST low--resolution spectral data, we are able to estimate the mass--accretion rates of T Tauri stars by analyzing various permitted emission lines of H$\alpha$ 6563~$\AA$, H$\beta$ 4861~$\AA$, Ca$\text{II}$ 8498, 8542, 8662~$\AA$, and He$\text{I}$ 4471, 5875, 6678, 7065~$\AA$ \citep[][]{2009A&A...504..461F, 2009ApJ...696.1589H, 2014A&A...561A...2A, 2017A&A...600A..20A}, which are generated by the accretion flows between the disc and star, in the available wavelength coverage.
The emission profiles of these lines in all spectra are shown in Figure~\ref{fig:figure_emission_profile-ha}, Figure~\ref{fig:figure_emission_profile-hb}, Figure~\ref{fig:figure_emission_profile-Ca}, and Figure~\ref{fig:figure_emission_profile-HeI}.
}

For each spectrum, we first estimate the equivalent width (EW) of the emissions with the equation:
\begin{equation}
EW = \int(\frac{F_{\lambda}}{F_{o}}-1)d\lambda,
\label{eq:eq1}
\end{equation}
where $F_{\lambda}$ is the spectral flux of the line defined to be within the 3--$\sigma$ of the best--fit Gaussian model with the central wavelength of the particular line; $F_{o}$ is the average flux of the continuum in the same range.
\textnormal{The equivalent widths of all available emissions in the spectra are listed in Table~\ref{tab:spectra-EW}.}
Then, the emission's flux ($F_{em}$) as well as their luminosities ($L_{em}$) can be obtained using the following equations:
\begin{equation}
F_{em} = EW_{em}F_{c}(em),
\label{eq:eq2}
\end{equation}
\begin{equation}
L_{em} = 4 \pi d^2 F_{em}.
\label{eq:eq3}
\end{equation}
Here, the $EW_{em}$ represents the equivalent width of the emission, $F_{c}(em)$ is the local continuum flux in the emission's wavelength range, and $d$ is the distance to the stars.
\textnormal{The mass--accretion luminosity ($L_{acc}$) of T Tauri star can be determined from the line luminosity ($L_{line}$) by using the relation below: 
\begin{equation}
log(L_{acc}/L_{\odot}) = b + a*log(L_{line}/L_{\odot}),
\label{eq:eq4}
\end{equation}
where $L_{\odot}$ is the solar luminosity. 
The coefficients $a$ and $b$ of every line used in the study is given by \cite{2017A&A...600A..20A} (see Tabel~B.1. in their paper).
}
If the accretion energy is reprocessed fully into the accretion continuum, the $L_{acc}$ can be converted into the mass--accretion rate ($\dot{M}_{acc}$) using the following equation from \cite{2008ApJ...681..594H}:
\begin{equation}
\dot{M}_{acc} = \frac{L_{acc}R_{*}}{GM_{*}( 1 - \frac{R_{*}}{R_{in}} )},
\label{eq:eq_mass_accretion_rate}
\end{equation}
where $M_{*}$ is the stellar mass, $G$ is the gravitational constant, $R_{*}$ is the stellar radius, and $R_{in}$ is the inner radius of the disk, which is defined as the disk truncation radius of CTTSs with a standard value of $5R_{*}$ \citep{1998ApJ...492..323G}.
\textnormal{The estimated mass--accretion rates from all available emissions are displayed in Table~\ref{tab:spectra-driven-Macc}.}

\subsection{Variability of our CTTSs sample}
A statistical metric system reported by \cite{2014AJ....147...82C} allows the light curves of young stellar objects to be classified into different categories, thus it is useful to recognize the burst--like accretion events in the TESS light curves of our samples. 
The system contains two metrics: the flux asymmetry (denoted as \emph{M}), and the quasi--periodicity (denoted as \emph{Q}).
The metric \emph{M} is the first one we focus here, which represents the light curve behavior tendency of dipping (\emph{M}~$>0$) and bursting (\emph{M}~$<0$).
\textnormal{We determine the metric \emph{M} following the approach explained in detail by \cite{2018AJ....156...71C} (see Sec. 5. in their paper).}
We first normalize the light curve with the median of the flux and smooth it with 200 points moving mean.
After subtracting the smoothed light curve from the original one, we identify the 5$\sigma$ outliers in the residual curve and remove their counterparts from the original curve.
Then the metric \emph{M} is estimated using the formula:
\begin{equation}
    M = - ( f_{5\%} - f_{med} )/\sigma_{f},
    \label{eq:metrix_M}
\end{equation}
where $f_{5\%}$ is the mean of the first and last 5 percent of the fluxes in the outlier--free curve, and $f_{med}$ and $\sigma_{f}$ are the median flux and the rms of the original curve, respectively.
Stars with \emph{M}~$<-0.25$ are likely to display the accretion--like burst variability in their light curves.
\textnormal{The \emph{M}--values of all of these CTTSs are listed in Table~\ref{tab:M+Q_value}.}

\begin{figure}
    \centering
    \epsscale{1}
    \plotone{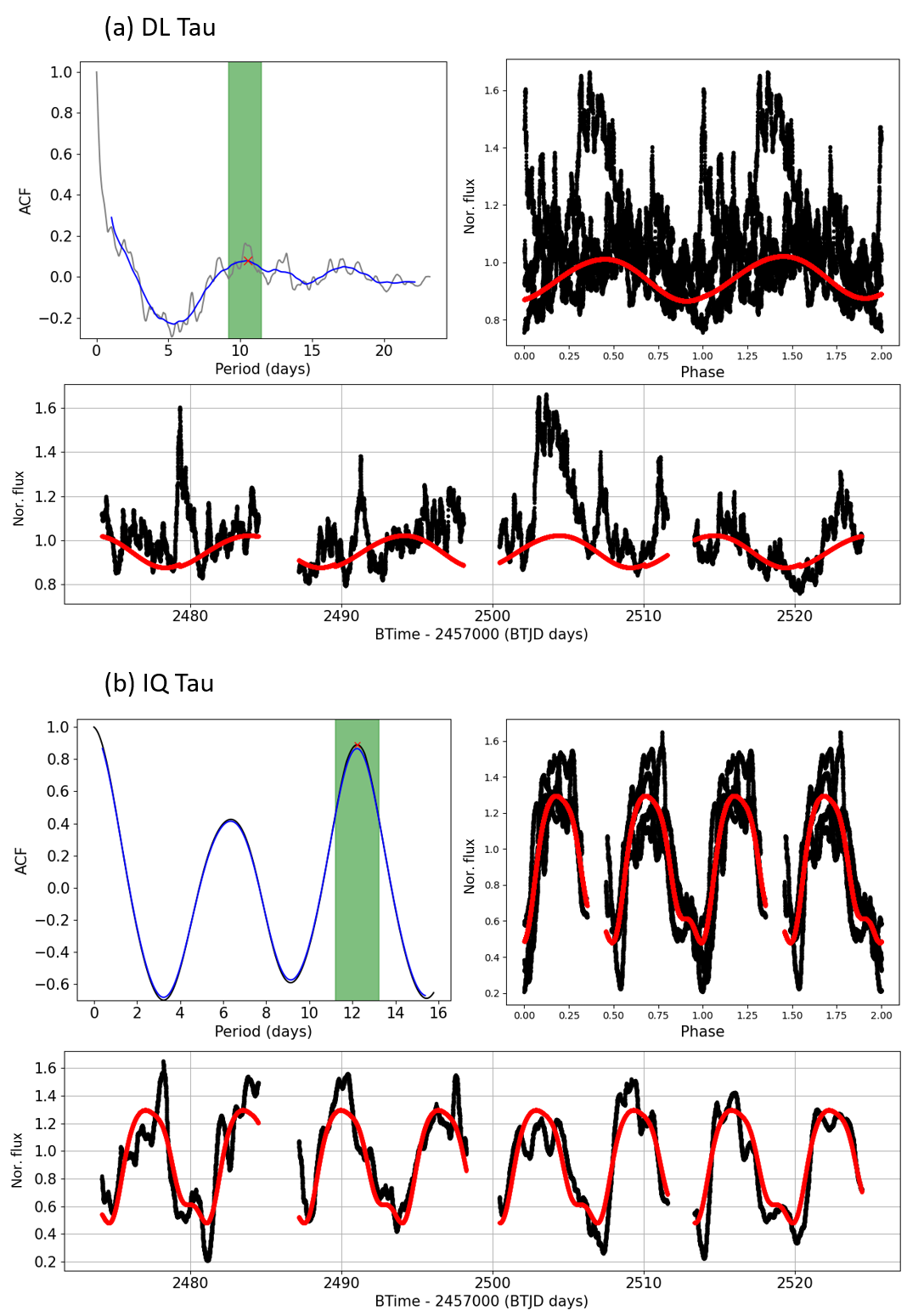}
    \caption{The examples of the period re-/estimate of the CTTSs, here shows (a) DL~Tau and (b) IQ~Tau. In each case, the upper left panel shows the autocorrelation function. The gray one is the original ACF, the blue one is a smoothed version of the ACF. The red cross marks the peak of the ACF, and the FHWM highlighted by green, from where the best period is determined.
    The upper right panel is the phase curve (black) and the best Gaussian Process fit curve (red).
    The panel at the bottom is the full TESS Section 43 and 44 light curve of the star (black) with the best Gaussian Process fit curve (red).
    The best periods we found for DL~Tau and IQ~Tau are 10.4 day and 12.9 day, respectively.
    }
    \label{fig:DL_Tau_as_example_period}
\end{figure}

\textnormal{The second metric \emph{Q} is used to evaluate periodicity tendency of light curves in the range from \emph{Q}=0 (purely periodic) to \emph{Q}=1 (stochastic) \citep{2014AJ....147...82C}.
To determine the Q--values of the light curve, we first fold up the light curves with the associated periods reported by \cite{2010PASP..122..753P} and \cite{2020AJ....159..273R} if any.
One special case is DQ~Tau, which is a spectrocopic binary and has two reported brightness variation periods: $\sim15.8$~day and $\sim3.03$~day.
The first one is attributed to the brightening effect associated with mass--accretion and magnetic reconnection that occurs around the periastron when two stars approach each other \citep{1997AJ....113.1841M, 2010A&A...521A..32S, 2016ApJ...818..156C, 2016AJ....152..188G}.
The second one is the rotational modulation caused by the presence of the spots, and we use it to determine the $Q$-value for DQ~Tau.
Since the morphologies of variability in young stellar objects could vary with time \citep[e.g.,][]{2016A&A...586A..47S}, we manually examine the phase curve to ensure the validity of the reported periods for our data.}
\textnormal{
If the star has no reported period, or if the reported period is not applicable to our data, we re--/determine the period by, firstly, identifying the maximum in the autocorrelation function and, secondly, refining the period within the full height width of the maximum (FHWM).
The best period solution is determined by selecting the one that provides the optimal folded phase curve counterpart. 
This selection process involves manually examining the phase curves and comparing them to those from other potential period values.
Figure~\ref{fig:DL_Tau_as_example_period} illustrates DL~Tau and IQ~Tau as examples, with estimated periods of 10.4 day and 12.9 day, respectively.}
\textnormal{Next, we use the Gaussian Process (GP) to produce a smooth light curve pattern with the optimal period, which is then subtracted from the original light curve to measure the residual noise. 
Finally, the resulting residual variance is divided by the variance of the original light curve to obtain the \emph{Q}-value.}

\begin{figure}
    \centering
    \epsscale{1.}
    \plotone{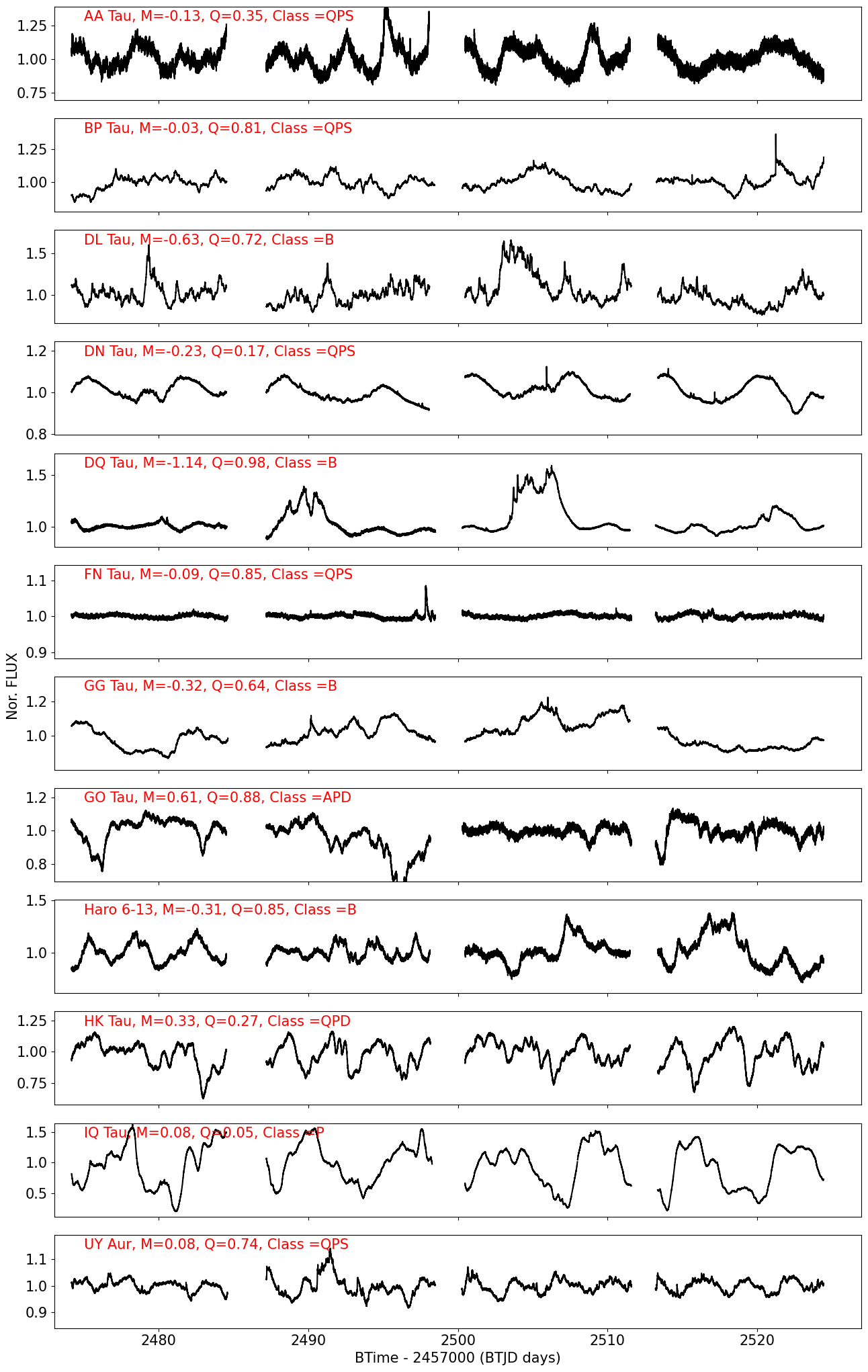}
    \caption{The TESS light curves of 12 CTTSs in our sample. Their variability classes are also displayed.}
    \label{fig:tess_M+Q_lightcurve}
\end{figure}

\begin{figure}
    \centering
    \epsscale{1.}
    \plotone{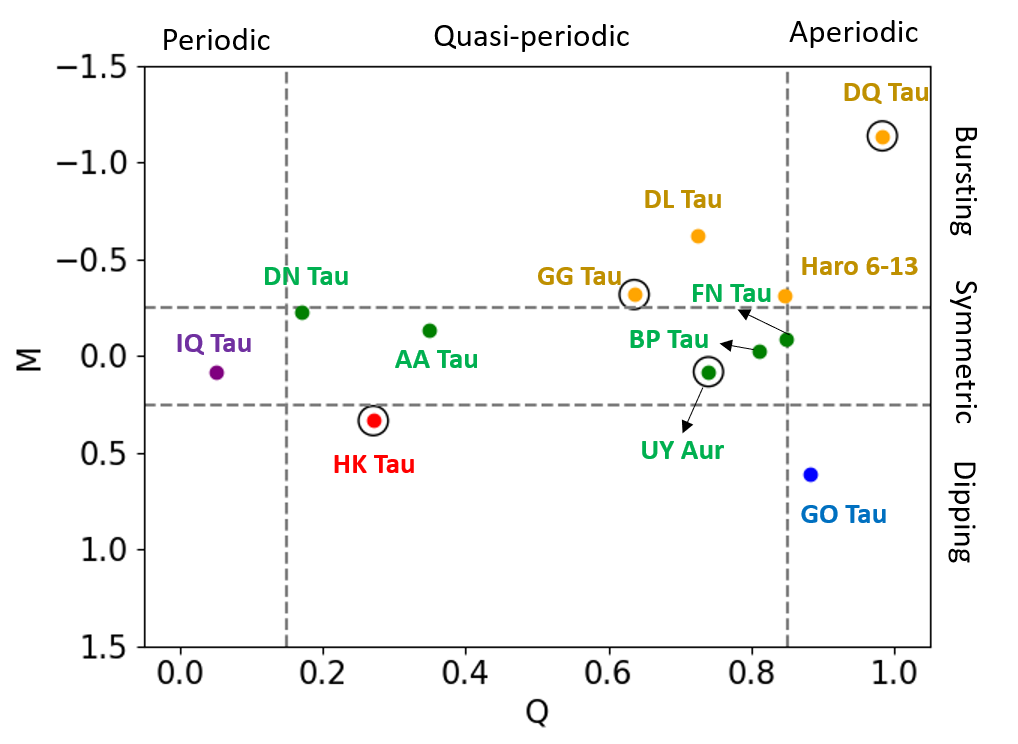}
    \caption{The $Q$--$M$ statistics diagram for our sample of CTTSs. The stars surrounded by black hollow circles in this diagram are binary or multi--star systems. Different colors represent different variability classes: yellow (B), purple (P), green (QPS), blue (APD), and red (QPD). }
    \label{fig:M_Q_space}
\end{figure}

\textnormal{We then divide the light curves of our samples into eight different categories using the definition of Q and M ranges given by \cite{2022AJ....163..212C}:
\begin{enumerate}
\item Burster (B): $M < -0.25$
\item Purely periodic symmetric (P): $Q < 0.15$ and $-0.25 < M < 0.25$
\item Quasiperiodic symmetric (QPS): $0.15 < Q < 0.85$ and $-0.25 < M < 0.25$
\item Purely stochastic (S): $Q > 0.85$ and $-0.25 < M < 0.25$
\item Quasiperiodic dipper (QPD): $0.15 < Q < 0.85$ and $M > 0.25$
\item Aperiodic dipper (APD): $Q > 0.85$ and $M > 0.25$
\item Long timescale (L)
\item Variable but unclassifiable (U)
\end{enumerate}
The values of \emph{Q} and \emph{M} for the CTTSs are listed in Table~\ref{tab:M+Q_value}, along with the variability classes.
The TESS light curves of 12 CTTSs in our sample, along with their assigned \emph{M}, \emph{Q} and variability classes, are shown in Figure~\ref{fig:tess_M+Q_lightcurve}.
Also in Figure~\ref{fig:M_Q_space}, we show the distribution of the variability statistics for our sample in the $Q$--$M$ diagram.}
\label{sec:sec3.3}

\subsection{Mass--accretion rate measurements from TESS light curves}

In Section~\ref{sec:sec3.3}, we have identified four bursters in our samples: DQ~Tau, DL~Tau, GG~Tau, and Haro~6--13.
To estimate the mass--accretion rates from their TESS light curves, we need to first convert the excess flux yields associated with the bursts in the $TESS$ bandpass to the U--band excess luminosity which can be used as the proxy for the mass--accretion rates \citep[e.g.,][]{2017ApJ...835....8T}.

\begin{figure}
    \centering
    \epsscale{1.2}
    \plotone{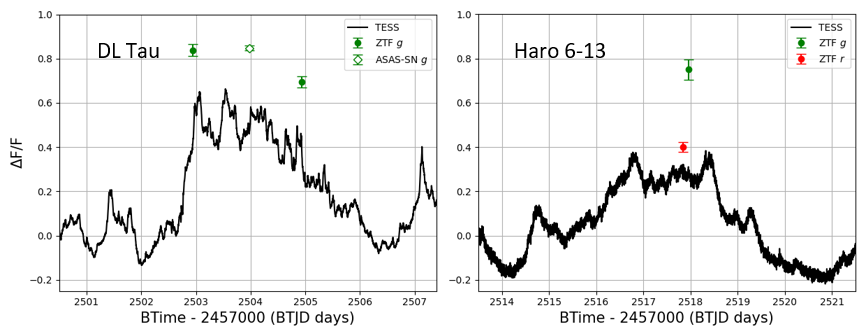}
    \caption{Left panel: A burst event of DL~Tau at BT-2457000 = 2504 was captured by TESS, ASAS--SN $g$, and ZTF $g$. Right panel: A burst event of Haro~6--13 at BT-2457000 = 2516 was observed by TESS, ZTF $g$, and ZTF $r$.
    In each panel, the black curve represents the burst event observed by TESS, and the green filled circles are $g$ band observations from the ZTF survey. The hollow green diamond represents the data from the ASAS--SN survey. In the case of Haro~6--13, the red filled circle represents the observation from the ZTF $r$ band.}
    \label{fig:DL_Tau_tess+asas-sn}
\end{figure}

\textnormal{To do so, we searched for other observational data of the stars from ASAS-SN, Pan--STARRS, and ZTF \citep{2019PASP..131a8002B, 2019PASP..131a8003M, ZTF2022}, that were taken simultaneously with TESS in different optical bands. 
We found such data only for DL~Tau and Haro~6--13.
DL~Tau was observed by the ASAS--SN survey and the ZTF survey. 
A burst event of DL~Tau at BT-2457000~=~2504~days was captured by TESS, ZTF $g$, and ASAS--SN $g$. 
Haro~6--13 was observed by ZTF during the same observation period as the TESS mission for it.
A burst event in Haro~--13's TESS light curve at BT-2457000~=~2516~days was also observed by the ZTF with both the $g$ and $r$ bands.
Their light curves of the bursts are shown together in Figure~\ref{fig:DL_Tau_tess+asas-sn}.}

\begin{figure}
    \centering
    \epsscale{0.7}
    \plotone{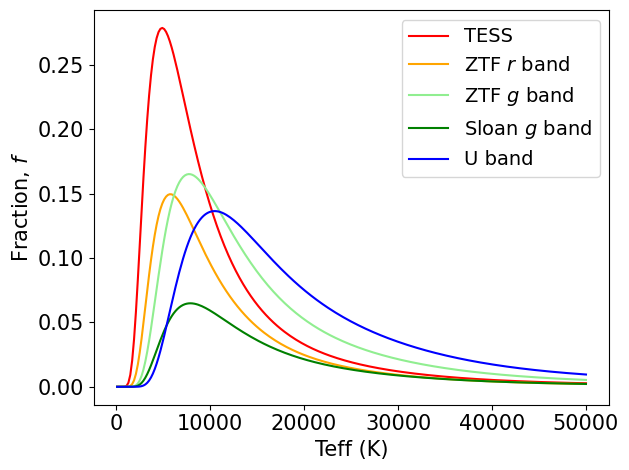}
    \caption{The fractional response functions of $TESS$ (red), ZTF~$r$ (orange), ZTF~$g$ (lightgreen), sdss~$g$ (green), and $U$ bands (blue). They are the function of the blackbody temperature.}
    \label{fig:band_fraction_response}
\end{figure}

Here, the periodic variation due to possible dark spot modulation has been \textnormal{removed} from the TESS light curves of DL~Tau and Haro~6--13 by subtracting the best--fit sinusoidal curves with corresponding periods (see Table~\ref{tab:M+Q_value})  from the original curves.
The fluxes of both data have good quality flags, and both are normalized to their respective median flux levels using the equation:
\begin{equation}
  \frac{\Delta F}{\widetilde{F}} = \frac{F-\widetilde{F}}{\widetilde{F}}, 
  \label{eq:eq_amplitude}
\end{equation}
\textnormal{where $F$ represents the flux as a function of time and $\widetilde{F}$ is their median in the light curve.}
We express the observed amplitude due to the hot spot associated with the accretion burst, that covers an area $A=X\pi R_{*}^{2}$ on the visible stellar photosphere, where $X$ is the filling factor and $R_{*}$ is the stellar radius as:
\begin{equation}
    \left(\frac{\Delta F}{\widetilde{F}}\right)_{\lambda} = X \frac{\int B_{\lambda}(T_{\text{hspot}})R_{\lambda}d\lambda}{\int B_{\lambda}(T_{*})R_{\lambda}d\lambda} = X \frac{T_{\text{hspot}}^4 f_{\lambda}(T_{\text{hspot}})}{T_{*}^4 f_{\lambda}(T_{*}) },
\label{eq:eq_burst_amp}
\end{equation}
where $B_{\lambda}$ is the Planck function, $R_{\lambda}$ is the transmission function of the band, $T_{*}$ is the stellar effective temperature, and $T_{hspot}$ is the temperature of the hot spot.
The integral term can be expressed as $\sigma T^{4} f_{\lambda}(T)$ where $\sigma$ is the Stefan--Boltzmann constant and $f_{\lambda}$ is the bandpass response factor as a function of temperature for the used band ($\lambda$).
\textnormal{The response factor functions of the bands used in the study, i.e., sdss~$g$, ZTF~$g$, ZTF~$r$, $U$, and $TESS$ bands, are shown in Figure~\ref{fig:band_fraction_response}.}
The amplitudes of DL~Tau's event are 0.848, 0.77, and 0.438 in ASAS-SN's sdss~$g$, ZTF~$g$, and $TESS$ bands, respectively.
\textnormal{We note that since there are two data points from ZTF~$g$ obtained during the event, the reported amplitude in ZTF~$g$ is computed as the mean value of these data points. Additionally, the central time between these two data points is very close to that of the ASAS-SN's observation.}
With this information, we can estimate the temperature of the hot spot by using the following equation:
\begin{equation}
\frac{(\Delta F/\widetilde{F})_{\lambda1}}{(\Delta F/\widetilde{F})_{\lambda2}} = \frac{f_{\lambda1}(T_{\text{hspot}})f_{\lambda2}(T_{*})}{f_{\lambda2}(T_{\text{hspot}})f_{\lambda1}(T_{*})}.
\label{eq:hotspot_Teff}
\end{equation}
\textnormal{The hot spot's temperatures derived from $TESS$--ASAS-SN~$g$ and $TESS$--ZTF~$g$ are about 5700~K and 5500~K, respectively.}
\textnormal{As a result, the temperature of the hot spot is estimated to be the average of these values, which is approximately 5600~K.
We substitute this value back to the Eq.~\ref{eq:eq_burst_amp} to calculate the filling factor of the spot and obtain $X=0.038$ or $3.8\%$.}

\textnormal{For the case of Haro~6--13, the burst amplitudes in TESS, ZTF~$r$, and ZTF~$g$ bands are measured to be 0.292, 0.4, and 0.75, respectively.
Because there is approximately a 0.1~day timing gap between the ZTF~$g$ and ZTF~$r$ data, the amplitude in the TESS band is the average of the corresponding amplitudes at ZTF~$g$ and ZTF~$r$ data times.
By following the same estimation procedure as we conducted for DL~Tau's burst, the temperature of Haro~6--13's burst is estimated to be 7100~K based on TESS and ZTF~$g$ bands data.
The estimated temperature becomes a bit hotter when using TESS and ZTF~$r$ bands data, which is 7715~K.
Thus we determined the burst temperature of Haro~6--13 to be about 7300~K by averaging these values.
The filling factor of the burst--driven hot spot on Haro~6--13 is about $X=0.008$ or $0.8\%$.}
The resulting values of the hot spot temperature and fractional area fall within the range of values observed and modeled in previous studies \citep[e.g.,][]{2013A&A...558A.114M,2015A&A...581A..66V}, demonstrating the reliability of our method.
However, we note that our estimations are lower limits as we haven't considered the limb--darkening in Eq.~\ref{eq:eq_burst_amp} and Eq.~\ref{eq:hotspot_Teff}.
In other words, we assume the hot spot is located at the center of the visible photospheric disk of the star.

\begin{figure}
    \centering
    \epsscale{1.2}
    \plotone{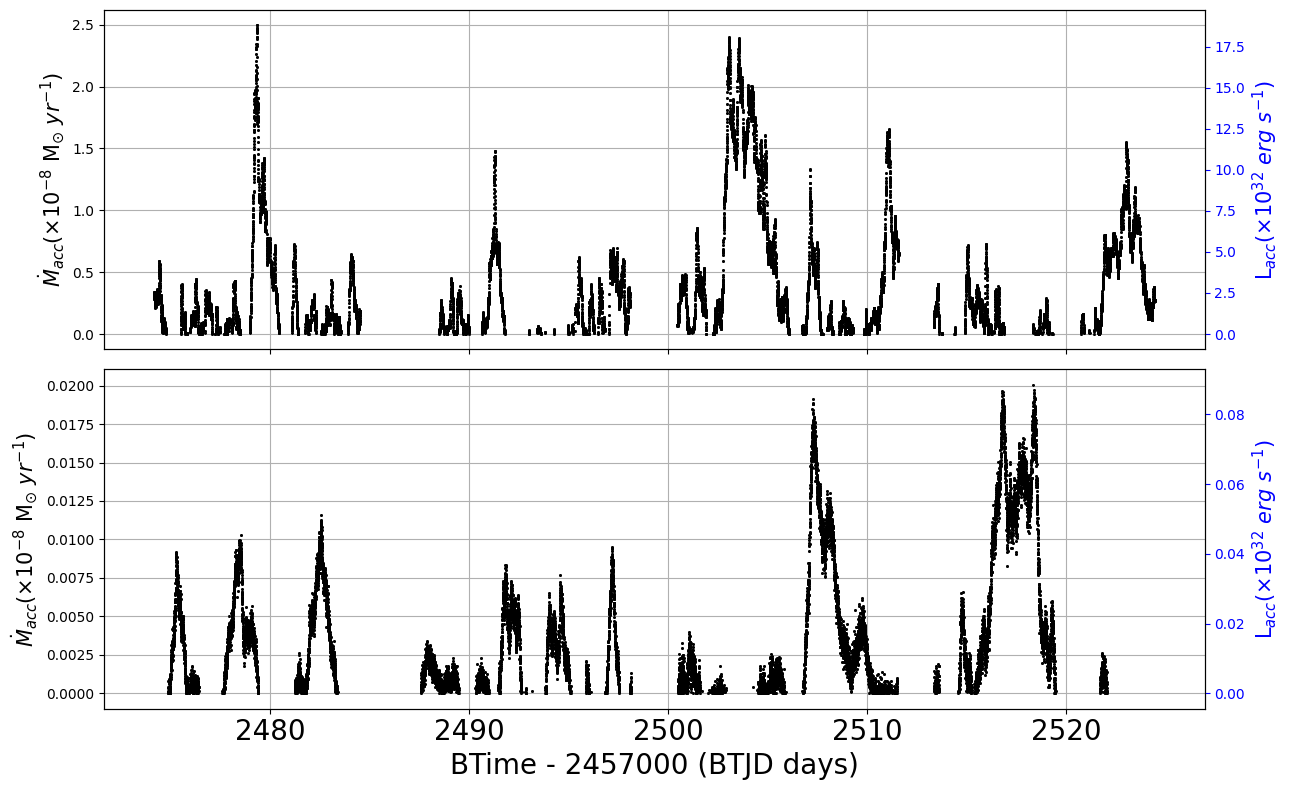}
    \caption{The time series mass--accretion rates of DL~Tau and Haro~6--13 are shown in the upper panel and bottom panel, respectively.
    The left axis shows the mass--accretion rates derived from the TESS, ASAS--SN, and ZTF observations, while the right axis displays the corresponding accretion luminosities.}
    \label{fig:DL_Tau_mass_accretion_rate}
\end{figure}

\textnormal{
We then extrapolate the burst amplitudes in the $U$--band with the estimated hot spot temperatures and the fractional areas.
We find that the burst amplitudes in the $U$--band are about 3.49 times and 7.6 times higher than that in the $TESS$ band for DL~Tau and Haro~6--13, respectively.
We produce the $U$--band excess amplitude curves of DL~Tau and Haro~6--13 by multiplying their normalized TESS light curves with the factors of counterparts.
}
\textnormal{
The U--band apparent magnitudes of DL~Tau and Haro~6--13 are $14.04$ \citep{2002yCat.2237....0D} and $19.22$ \citep{2007A&A...468..379A}, which can be converted to a flux unit of $F_{U}=3.77\times10^{-11} erg~s^{-1} cm^{-2}$ and $F_{U}=3.19\times10^{-13} erg~s^{-1} cm^{-2}$, respectively.
The U--band luminosities, $L_{U} = 1.15\times10^{32} erg~s^{-1}$ and $L_{U} = 6.32\times10^{29} erg~s^{-1}$, of DL~Tau and Haro~6--13 can be estimated using the formula $L_{U}=F_{U}4\pi d^2$ with the distance information of $d=159$~pc and $d=129$~pc \citep{2020yCat.1350....0G}.
We produce the U--band excess luminosity ($L_{Uexcess}$) curves of these stars by multiplying their U--band excess amplitude curves with the $L_{U}$ values.
Then the accretion luminosities ($L_{acc}$) can be calculated using an empirical correlation between $L_{acc}$ and $L_{Uexcess}$ given by \cite{1998ApJ...492..323G}:
\begin{equation}
    log(L_{acc}/L_{\odot}) = 1.09log(L_{Uexcess}/L_{\odot}) + 0.98,
    \label{eq:eq8}
\end{equation}
where $L_{\odot}$ is solar luminosity.
Finally, the mass--accretion rates can be determined from the accretion luminosities by using Eq~\ref{eq:eq_mass_accretion_rate} with the stellar parameters in Table~\ref{tab:tab2}.
The mass--accretion rates and luminosities of DL~Tau and Haro~6--13 are displayed in Figure~\ref{fig:DL_Tau_mass_accretion_rate}.
}

\section{Results and Discussion}
\label{sec:section4}

\subsection{Mass--accretion rates from the LAMOST Spectra}
\textnormal{We estimated the mass-accretion rates of low-mass classical T Tauri stars with spectral types from mid-M to late-K in the Taurus Molecular Cloud using LAMOST spectral data. The indicators we used were the H$\alpha$ 6563~$\AA$, H$\beta$ 4861~$\AA$, Ca$\text{II}$ 8498, 8542, 8662~$\AA$, and He$\text{I}$ 4471, 5875, 6678, 7065~$\AA$ emissions. The equivalent widths of these lines and the corresponding mass-accretion rates are listed in Table~\ref{tab:spectra-EW} and Table~\ref{tab:spectra-driven-Macc}, respectively.
}

\begin{figure}
    \centering
    \epsscale{1.2}
    \plotone{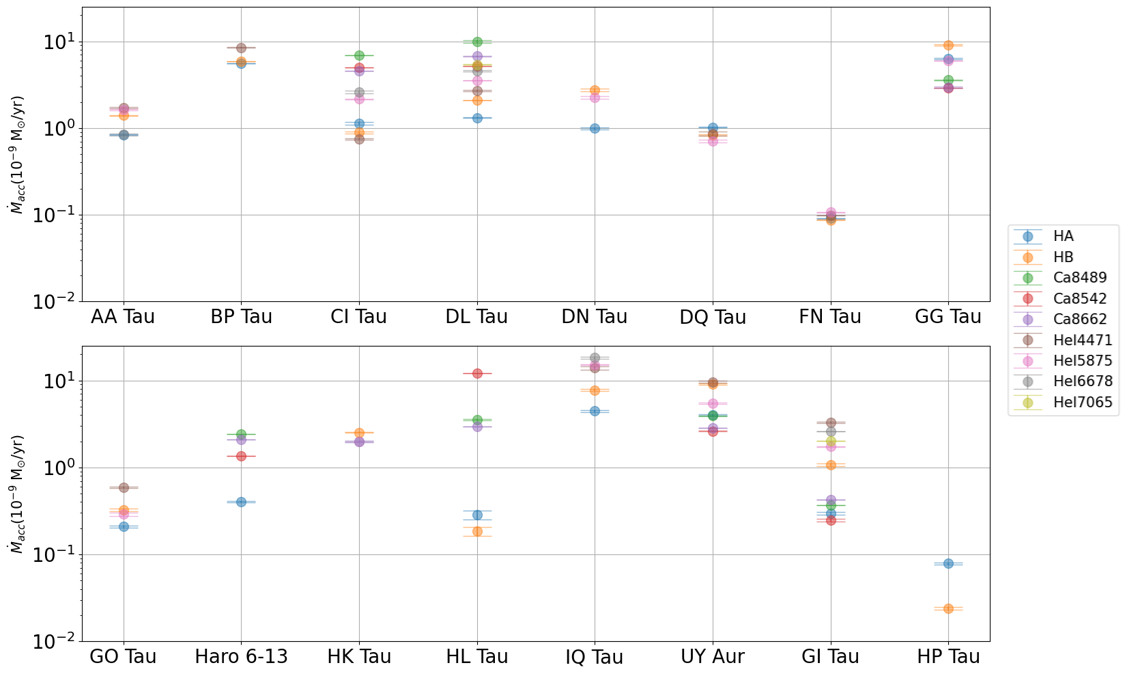}
    \caption{The distribution of mass-accretion rates derived from all available emissions in the spectra of our CTTSs. The names of the stars marked with an asterisk (*) in the upper right indicate that they are binary or multi--star systems.}
    \label{fig:spectra_Macc}
\end{figure}

\textnormal{
All of our samples have the $\dot{M}_{acc}$ measurement estimated from H$\alpha$ line.
On average, these stars have an H$\alpha$ mass--accretion rate of $1.76\times10^{-9} M_{\odot}yr^{-1}$. 
GG~Tau has the highest measured H$\alpha$ mass--accretion rate among the sample, with a value of $6.35\times10^{-9} M_{\odot}yr^{-1}$.
HP~Tau has a significant H$\alpha$ emission, but its mass--accretion rate is lower than those of the other stars with a value of $7.88\times10^{-11} M_{\odot}yr^{-1}$ in this study.}
\textnormal{
On the other hand, from H$\beta$ emissions, the average rate of these star is $2.89\times10^{-9} M_{\odot}yr^{-1}$.
UY~Aur and GG~Tau both have the highest H$\beta$--derived rate of $9.11\times10^{-9} M_{\odot}yr^{-1}$.
Consistent with the H$\alpha$ result, HP~Tau exhibits the lowest mass--accretion rate estimated from the H$\beta$ line, which is $2.44\times10^{-11} M_{\odot}yr^{-1}$.
Only Haro~6--13 has no H$\beta$ accretion rate because it is too faint for LAMOST in the blue wavelength part, making it impossible to estimate.
}
\textnormal{
Not all targets in our sample have measurable equivalent widths for the other emission lines, which means that not all corresponding mass--accretion rates could be estimated. 
Figure~\ref{fig:spectra_Macc} shows the mass-accretion rate distribution of our targets by displaying the measurements derived from all available emissions adopted in this study.
HL~Tau has the largest dispersion of the spectral-derived mass-accretion rates with a coefficient of variation (CV) value (the ratio of the standard deviation to the mean) of $1.15$, as compared to other stars in the sample.
The CVs of other stars are smaller than one with a mean value of 0.41$\pm$0.22.
After field checking, we have ruled out the possibility of contamination from the closest neighboring star of HL~Tau, XZ~Tau, as it is located about $23''$ away from HL~Tau, and thus, can be resolved separately by the LAMOST fiber with a $3.3''$ diameter. 
The underlying mechanism responsible for this phenomenon remains unknown, making it an intriguing topic for future investigation.
On the other hand, the lowest dispersion CV value is found in FN~Tau, which is $0.08$.
}


\subsection{Mass--accretion rates of DL~Tau and Haro~6--13 from the TESS light curves}
The time--series mass accretions rates and luminosities derived from the TESS light curves for DL~Tau and Haro~6--13 are shown in Figure~\ref{fig:DL_Tau_mass_accretion_rate}.
\textnormal{We find that the mass--accretion rate of DL~Tau reached up to $2.5\times10^{-8} M_{\odot}yr^{-1}$ during the observation time. 
The mass accreted by DL Tau over approximately 50 days is $\approx 3.4\times10^{-10} M_{\odot}$, with an average rate of $\approx 2.5\times10^{-9} M_{\odot}yr^{-1}$.} 
These results are comparable to the mean value of the mass--accretion derived from LAMOST for DL~Tau, which is $4.6\times10^{-9} M_{\odot}yr^{-1}$ with a standard deviation of $2.5\times10^{-9} M_{\odot}yr^{-1}$.
\textnormal{Meanwhile, the maximum mass--accretion rate of Haro~6--13 from the TESS observations is $2\times10^{-10} M_{\odot}yr^{-1}$.
It accreted mass with a value of about $2.7\times10^{-12} M_{\odot}$ over approximately 50 days, with an average rate of $\approx 2\times10^{-11} M_{\odot}yr^{-1}$.
These results are overall lower than the measurement, which is about $1.6\pm0.8\times10^{-9} M_{\odot}yr^{-1}$ on average, determined from LAMOST data.}


An interesting question is whether the mass increase of T Tauri stars is dominated by long--term, steady mass accretion or by occasional bursts of accretion? 
First, we need to define a steady accretion rate, and this rate does not change over time. 
We then calculate the total mass accumulated by the steady rate of the star during the TESS observation.
This is compared with the total amount of burst accretion to determine whether the 
CTTS is dominated by steady status or episodic effect.


\textnormal{Here, we take the mean rate obtained from LAMOST emission measurement as the steady rate. 
Thus, the steady rate of mass--accretion of DL~Tau is $4.6\pm2.5 \times10^{-9} M_{\odot}yr^{-1}$.
In this case, DL~Tau accumulated masses of $6.3\pm3.4 \times10^{-9} M_{\odot}$ through steady rate over a period of about 50 days.
This result is larger than the total mass of $\approx 3.4\times10^{-10} M_{\odot}$ accreted by the stars' burst--like events observed by TESS.
Similarly, the steady mass--accretion of Haro~6--13 is $1.6\pm0.8\times10^{-9} M_{\odot}yr^{-1}$.
This value is about one order higher than the maximum of $2\times10^{-10} M_{\odot}yr^{-1}$ from the TESS observation. 
Therefore, in this scenario, the mass--accretion processes of DL~Tau and Haro~6--13 are both steady accretion dominant.
}

\subsection{\textnormal{Flare activity of our CTTSs sample}}
\textnormal{CTTSs are also known for their high magnetic activity, which is often manifested in flaring events.
Flares with the sudden, short--lived increases in brightness are thought to be caused by magnetic reconnection events on the star's surface \citep{1963ApJS....8..177P, 2012Natur.485..478M}. 
The magnetic fields of the CTTS form a large loop connecting the star and circumstellar disk. 
When the magnetic field lines eventually reconnect, a large amount of energy is released in the form of a huge flare.
This kind of the flare resulting from the star--disk magnetic interaction could last more than 50,000 seconds, and it is a characteristic feature of CTTS while affecting the mass--accretion process in some degree \citep[][]{2007A&A...463..275G, 2016A&A...590A...7L, 2019A&A...624A..50C}.
}
\subsubsection{\textnormal{Flare detection}}
\begin{figure}
    \centering
    \epsscale{1.2}
    \plotone{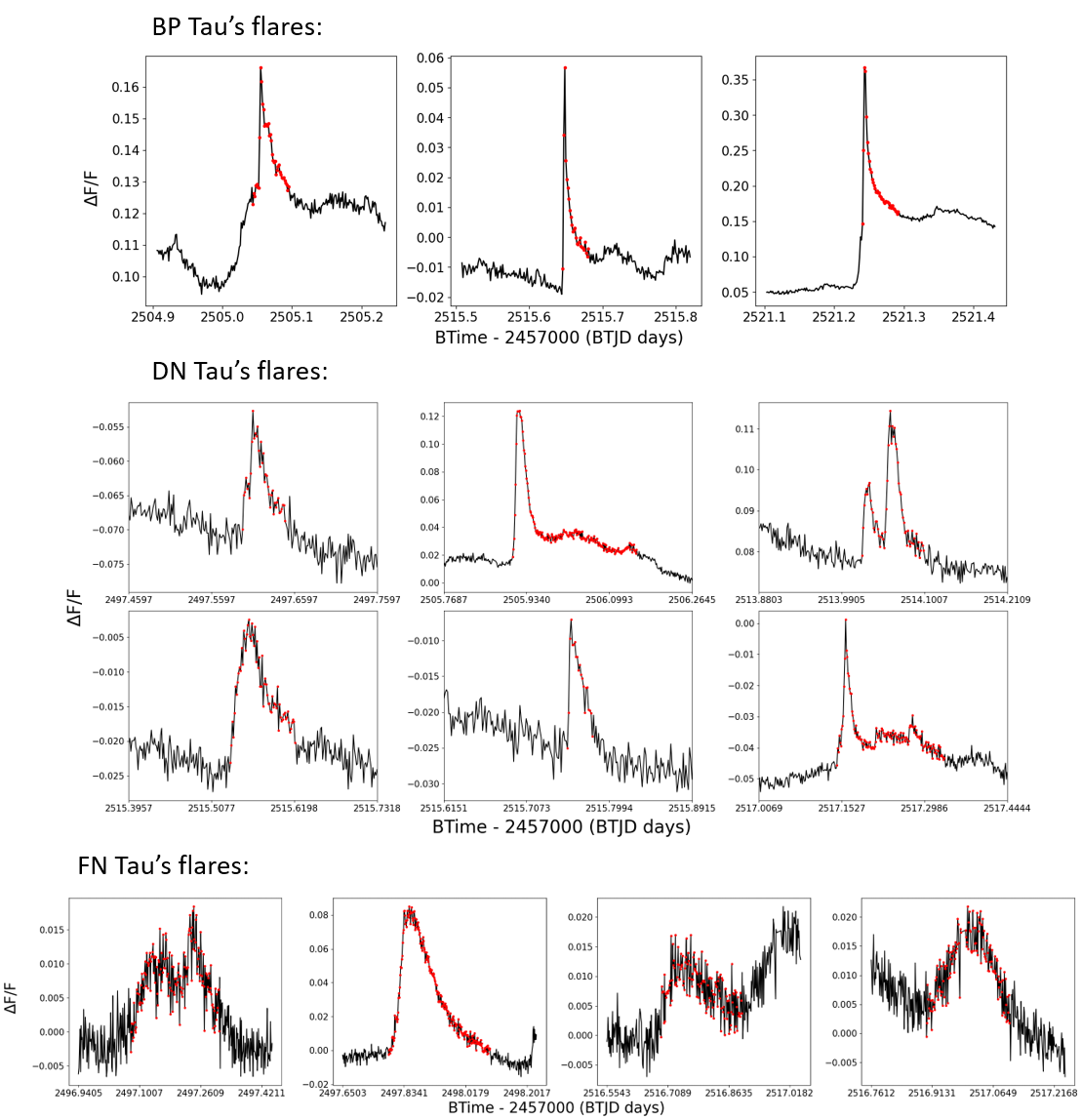}
    \caption{The profiles of 13 flare events detected in the Section 43 and 44 TESS light curves of BP~Tau (3 flares), DN~Tau (6 flares), and FN~Tau (4 flares).}
    \label{fig:13_flares}
\end{figure}

\textnormal{To analyze the flares in our TESS light curves, we employ an algorithm developed by \cite{2019ApJ...873...97L} that generates a flare--free light curve. 
We then identify individual flare candidates from a residual curve generated by subtracting the flare--free light curve from the original curve. 
This method allows us to accurately quantify the properties of each flare, including its amplitude, duration, and energy release.
We further examine the candidates for their authenticity by visually inspecting the Target Pixel File to generate the light curve of each pixel, using the Python package Lightkurve \citep[][]{2018ascl.soft12013L} and the method described by \cite{2015ApJ...798...92W}.
In our sample, we detect flares in the light curves of three stars: BP~Tau (3 flares), DN~Tau (6 flares), and FN~Tau (4 flares), in total 13 flares, and some of them are multi--flare events.
The light curve profiles of these flares are shown in Figure~\ref{fig:13_flares}.
However, the absence of flares in other stars does not necessarily mean that they are not flaring stars. 
For example, \cite{2017ApJ...835....8T} and \cite{2018ApJ...862...44K} both detected several flare events in DQ~Tau from optical and Kepler observations, whereas we did not detect any flares in DQ~Tau from the TESS observation in this study.
It is possible that there are microflares with amplitudes or duration that are too small to be detected by TESS.
}
\subsubsection{\textnormal{Flare amplitude and energy}}
The amplitude of the flare profile as a function of time can be determined using the Eq.~\ref{eq:eq_amplitude}.
The flare energy is estimated using an equivalent duration (ED) method defined as follows \citep{Gershberg1972}:
\begin{equation}
\mathnormal{
  ED = \int \frac{\Delta F(t)}{\widetilde{F}} dt . }
  \label{eq:ED}
\end{equation}
It represents the time it takes for a non-flaring star to emit the same amount of energy as that released during a given flare.
Therefore, this quantity can be used to estimate the energy of the flare by multiplying it by the quiescent stellar flux in the telescope bandpass
\begin{equation}
\mathnormal{
  E_{f, \lambda} = ED \times F_{*, \lambda}, }
  \label{eq:ED_flare_energy}
\end{equation}
and the corresponding quiescent stellar flux of the star can be estimated from
\begin{equation}
\mathnormal{
  F_{*, \lambda} = 4 \pi R_{*}^{2} \sigma T_{*}^{4} f_{\lambda}(T_{*}). }
  \label{eq:quiescentflux_for_flare}
\end{equation}
Here, $R_{*}$ is the stellar radius, $T_{*}$ is the effective temperature, and $\sigma$ is the Stefan--Boltzmann constant.
\textnormal{$f_{\lambda}$ is the same function as appeared firstly in Eq.~\ref{eq:eq_burst_amp}, which is a bandpass response factor as a function of temperature for the wavelength ($\lambda$) band and in this case $\lambda=TESS$, i.e., the fractional response function of TESS (see Figure~\ref{fig:band_fraction_response}).}

\textnormal{The amplitudes, duration, timing at peak in BTJD, and released energy of detected flares are listed in Table~\ref{tab:flares}. 
These flares have energy levels ranging from $2\times10^{34}$~erg to $6\times10^{35}~$erg. 
The flare with the highest amplitude of 0.22 is observed in BP~Tau, with a duration of about 76 minutes and energy of $1\times10^{35}$erg. 
The most energetic flare in our sample is observed in FN~Tau, with an energy of $6.3\times10^{35}$~erg, an amplitude of 0.09, and a duration of 438 minutes.
Furthermore, given the duration of the flares, these events likely originate from the localized magnetic reconnections near the stellar surface.}

\subsubsection{\textnormal{Flare frequency distribution}}
\begin{figure}
    \centering
    \epsscale{1.1}
    \plotone{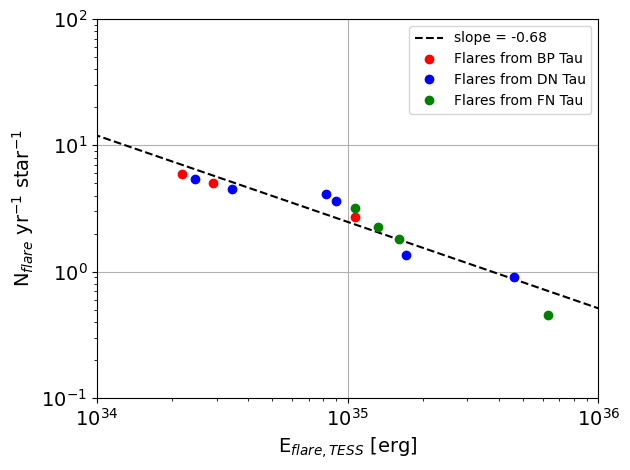}
    \caption{The flare frequency distribution and the best--fits power--law function with a slope of -0.68$\pm$0.20 of our CTTS sample based on 13 detected flares in this study.}
    \label{fig:flare_frequency_distribution}
\end{figure}

\textnormal{The flare activity of a star} can be characterized by its flare frequency distribution, which can be described by a linear relationship expressed in logarithmic form as
\begin{equation}
    \mathnormal{
    log~N=a+\beta~log~E,
    }
    \label{eq:eq11}
\end{equation}
where $N$ is the cumulative frequency of flares with released energy $E$. The slope of the distribution, represented by the coefficient $\beta$, can provide information about the properties of the star's flare activity. If $\beta <$-1, then most of the total energy emitted by the flares comes from small flares, while for $\beta >$-1, stronger flares dominate the energy output. 
The flare frequency distribution power--law can also be expressed in differential form as
\begin{equation}
    \mathnormal{
    dN \propto E^{-\alpha}dE,
    }
    \label{eq:eq12}
\end{equation}
where $dN$ is the number of flares with energies in the range of $E$ and $E+dE$, and $\alpha = 1 - \beta$ \citep{hawley2014kepler}.
We use the maximum likelihood estimator (MLE) for $\alpha$ as derived by \cite{Clauset2009}:
\begin{equation}
    \mathnormal{
    \alpha = 1 + n \left[ \sum^{n}_{i=1} \ln\frac{E_{i}}{E_{min}}  \right]^{-1},
    }
    \label{eq:eq13}
\end{equation}
and the error for $\alpha$ can be estimated from
\begin{equation}
    \mathnormal{
    \sigma_{\alpha} = \frac{\sqrt{n+1}(\alpha-1)}{n}.
    }
    \label{eq:eq14}
\end{equation}
Here, $n$ is the number of flare samples, and $E_{min}$ is the minimum value of the energy of the observed flares.
By setting $E_{min}=2.2\times10^{34}$~erg and $n=13$, we have calculated the value of $\alpha$ to be $1.68\pm0.20$. 
From this, we can derive the linear slope of logarithmic flare frequency distribution, which is represented by $\beta=-0.68\pm0.20$ and a constant $a$ of $24.32$.
Figure~\ref{fig:flare_frequency_distribution} shows the flare frequency distribution and its best--fit power--law slope for our CTTS sample. 
These results suggest that strong flares with high energy dominate the flare activity of these stars. 
The power--law distribution also indicates that our sample stars exhibit extreme flare activity, with the ability to generate flares with energy levels stronger than $E=1\times10^{34}$~erg about 12 times per year.
This is in stark contrast to solar-type stars, which produce flares of this magnitude only once every 35--40 years \citep{Shibayama2013}.
Even one of the most flare--active young M stars, Wolf~359, can produce only one such flare a year \citep[][]{2021AJ....162...11L}.

\subsection{The changes of the variability classes}
The changes in the variability behavior of the CTTSs can be attributed as the inner disk structure responsible for the stellar occultation, which can undergo significant changes in just a few years, transitioning from a stable and well-organized geometry with a consistent inner disk warp to a more disordered distribution of dust \citep[e.g.,][]{2015A&A...577A..11M}. 
Another explanation is the stability of the accretion regime. 
Initially, the accretion geometry may consist of a main accretion funnel in each hemisphere, with the base of each funnel corresponding to the stable inner disk warp. 
However, over time, this geometry may become unstable, resulting in the formation of random accretion funnels \citep[e.g.,][]{2013MNRAS.431.2673K}.
Cold spots due to the strong magnetic field of CTTSs might be able to cover a \textnormal{large fraction of stellar surface, which is similar to magnetically active} low--mass stars \citep[e.g.,][]{yang2017flaring, 2021AJ....162...11L}, and the changes of their distribution and sizes may be also reflected in the changes of variability behaviors \citep[e.g.,][]{2005A&A...432..647F}.

Ten stars in our sample were previously observed by K2~Campagin~13 (AA~Tau, DQ~Tau, GO~Tau, Haro~6--13 HK~Tau, HL~Tau, and IQ~Tau), or TESS Section 19 (BP~Tau, FN~Tau, and UY~Aur), and their light curves were analyzed and classified for variability by \cite{2022AJ....163..212C} and \cite{2022ApJ...935...54R}.
The variability classes given by \cite{2022AJ....163..212C} and \cite{2022ApJ...935...54R} are also listed in Table~\ref{tab:M+Q_value}. 
\textnormal{By comparing the variability classes given by \cite{2022AJ....163..212C} based on the K2~C13 data observed about four years ago, we find that most of the stars (6 out of 7) change their variability classes except DQ~Tau.
Specifically, AA~Tau changes from "S" to "QPS", GO~Tau changes from "U" to "APD", Haro~6--13 changes from "S" to "B", HK~Tau changes from "QPS" to "QPD", IQ~Tau changes from "QPS" to "P", and HL~Tau changes from "QPS" to "QPD".}
In contrast, the variability classes of BP~Tau and FN~Tau do not change compared to those derived by \cite{2022ApJ...935...54R} using TESS Section~19 light curves from about 600 days ago. 
Even though UY~Aur changes from "B" to "QPS" in 600 days, we still see a burst--like event at BJTD~$=$~2491 in its TESS Section 43 light curve.
These findings may suggest that the long--term timescale of the variation associated with, for example, the stable/unstable accretion regime and inner disc warp due to the star--disk interaction of the CTTSs is in the range between about 1.6 years to 4 years or longer.
However, it remains to be determined whether these changes occur in a cyclical or random manner. 
For instance, Haro~6--13 changing from the "B" type back to the "S" type, or to any other type, would require further investigation through continuous monitoring in the future.

\subsection{\textnormal{Effect of the inclination on the spectra, accretion, and variability properties}}
The spectral properties of the CTTSs have been predicted to be correlated with the inclination of the system, especially with the broad or high--velocity component of the forbidden line profile \citep[e.g.,][]{1995ApJ...452..736H}.
The high-velocity components are generally believed to originate in fast outflows that keep thrusting material into interstellar space in form of jets.
Among the known forbidden lines, the ones with the most distinctive high--velocity component spectral features are the [NII] lines at 6548~$\AA$ and 6583~$\AA$, which behave as blueshifted emission with a well-defined peak \citep{1997A&AS..126..437H}.
Nevertheless, the resolving power of the LAMOST observations is not high enough to distinguish the [NII] lines from H$\alpha$ emission, thus we can not derive the outflow velocity from these lines using the data we have in this study.

\textnormal{Additional forbidden lines that can provide insight into the wind velocity observed by LAMOST are the [OI] line at 6300~$\AA$, [SII] lines at 6716~$\AA$ and 6731~$\AA$, and [CaII] lines at 7291~$\AA$ and 7323~$\AA$. 
While the blueshifted emission of the [OI] line may not display a distinct peak when compared to the [NII], [SII], and [CaII] lines, the extended blue wing can be interpreted as an indicator of the projected wind velocity \citep[][]{2013A&A...558A..83A}.
Out of the sample CTTSs, ten stars exhibit [OI] lines in their LAMOST spectra, while [SII] lines have only been detected in AA~Tau and HL~Tau. 
Moreover, [CaII] lines are observed in only one star, HL~Tau.
}

\begin{figure}
    \centering
    \epsscale{1.2}
    \plotone{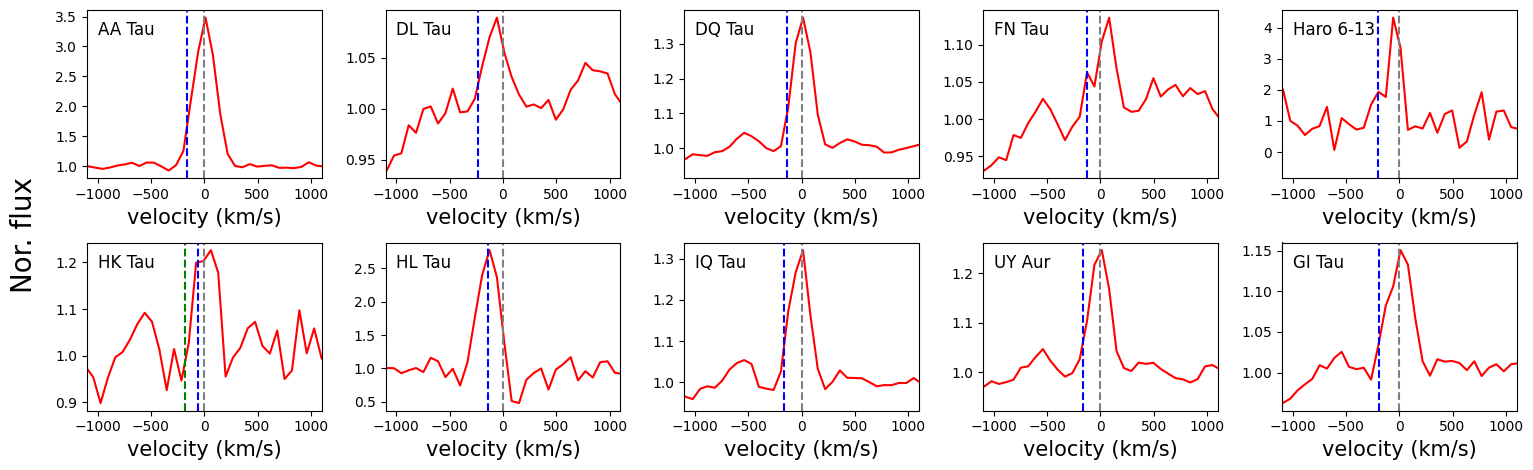}
    \caption{The velocity profiles of [OI]~6300 lines of ten CTTSs in our sample. In each panel, the gray vertical dashed line indicates the center of the line, and blue dashed line marks the wind velocity of the blueshifted wind or peak of the line. In a special case of HK~Tau, we also show a green dashed line to represent the blueshifted wind velocity, which is likely originating from a companion of the binary system, HK~Tau~A, and the blueshifted peak velocity marked with a blue dashed line belongs to the other companion of the system, HK~Tau~B.}
    \label{fig:QI6300_profiles}
\end{figure}

\begin{figure}
    \centering
    \epsscale{0.7}
    \plotone{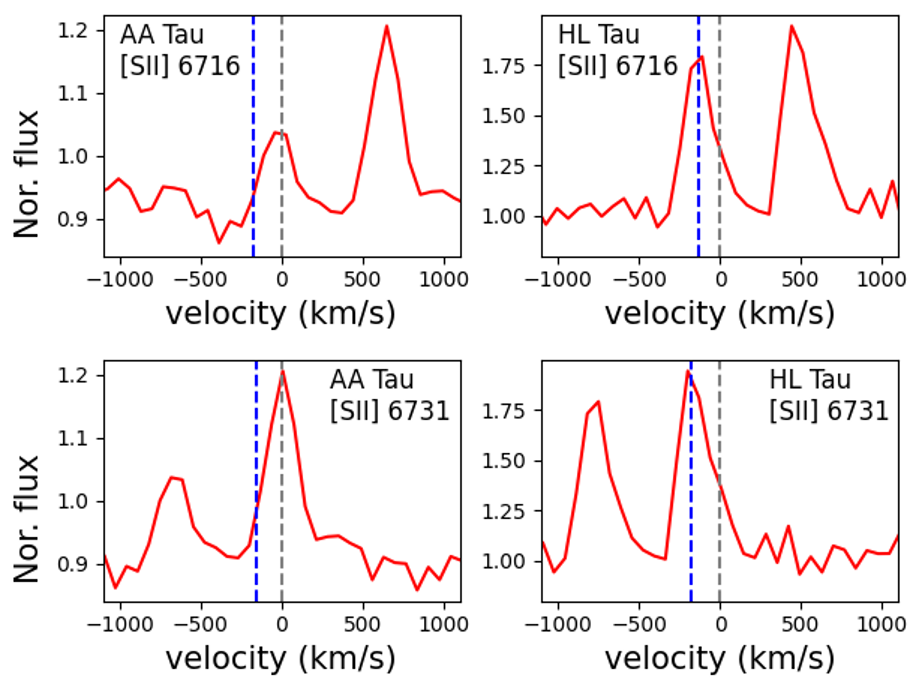}
    \caption{The velocity profiles of [SII]~lines observed in the spectra of AA~Tau and HL~Tau. In each panel, the gray vertical dashed line indicates the center of the line, and blue dashed line marks the wind velocity of the blueshifted wind or peak of the line.}
    \label{fig:SII_profile}
\end{figure}

\begin{figure}
    \centering
    \epsscale{0.6}
    \plotone{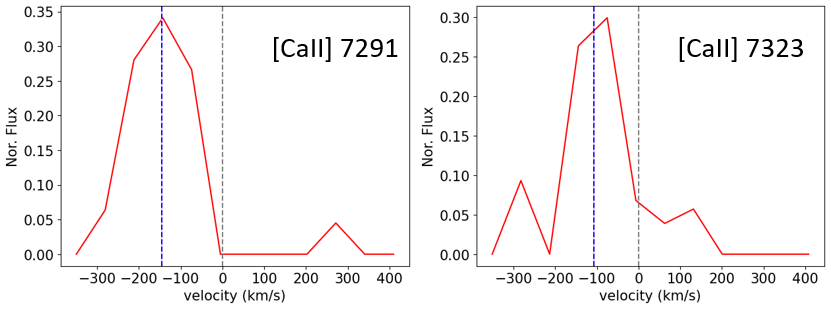}
    \caption{The velocity profiles of [CaII]~lines of HL~Tau. The gray vertical dashed line indicates the center of the line, and blue dashed line marks the wind velocity of the blueshifted peak of the line. }
    \label{fig:CaII_profile}
\end{figure}

\begin{figure}
    \centering
    \epsscale{1}
    \plotone{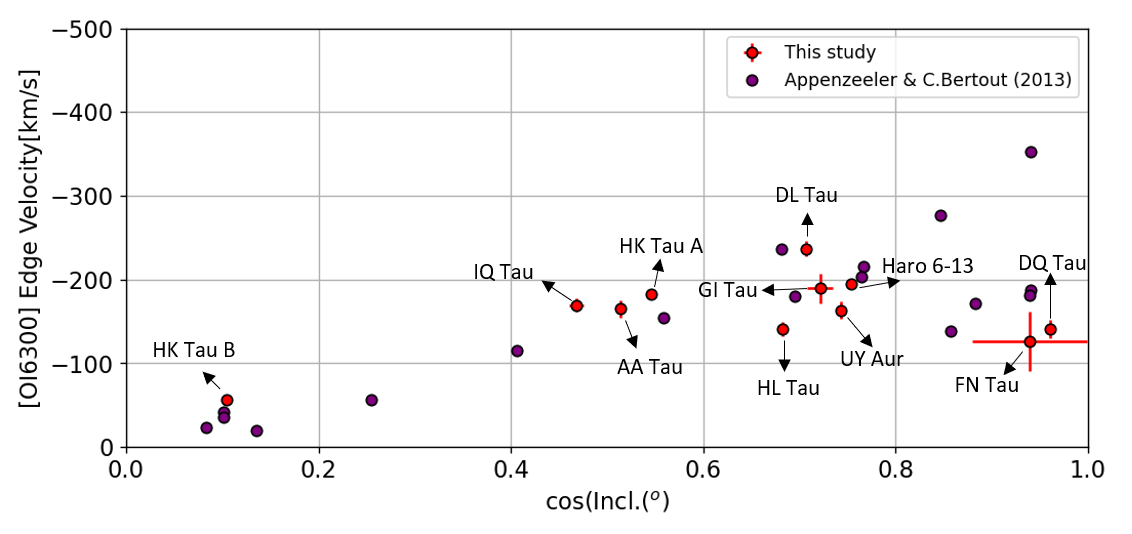}
    \caption{The comparison between CTTSs' OI~6300 project wind velocities (vertical axis) and inclinations in terms of cosine angle (horizontal axis). 
    The purple circles represent the data from \cite{2013A&A...558A..83A}, and the red circles are the data in this study.}
    \label{fig:QI_velocity_vs_inclination}
\end{figure}

\textnormal{
We estimate the projected wind velocity by using the blue edge of the line profile with flux reaching 25$\%$ of the peak value when the line shows no clear blue peak.
Because of the low resolution of the LAMOST data, for those profiles showing a blue peak, the velocity is directly estimated from the wavelength of the peak. 
The LAMOST [OI] line, [SII] lines, and [CaII] lines velocity profiles of the CTTSs are shown in Figure~\ref{fig:QI6300_profiles}, Figure~\ref{fig:SII_profile}, and Figure~\ref{fig:CaII_profile}, respectively.
The results of [OI], [SII], and [CaII] velocity measurements are listed in Table~\ref{tab:OI6300_velocity}. 
For stars where velocity measurements are available from not only [OI], but also [SII] and [CaII], i.e., AA~Tau and HL~Tau, the results are consistent across all lines. 
We note that there are two [OI] velocity results for a binary HK~Tau. 
The diameter of LAMOST's fiber is about 3.3 arcsecond, and the angular separation between HK~Tau~A and HK~Tau~B is about 2.3 arcsecond.
Thus, LAMOST observation can not resolve these two stars separately.
We detect a blueshifted peak of [OI] line at a wind velocity of $-56.49$~km/s, which may have originated from the edge--on system HK~Tau~B with an inclination of $84^{o}$. 
The velocity derived from the blue edge of the line with a value of $-182.88$~km/s belongs to HK~Tau~A, which has a relatively face--on disk with an inclination of $i=56.9^{o}$.
The comparison between the inclinations and [OI] line-derived projected wind velocity is shown in Figure~\ref{fig:QI_velocity_vs_inclination}.
Our results show a positive correlation between the cosine of inclination and the projected wind velocity, consistent with the trend observed by \cite{2013A&A...558A..83A}.
}

\begin{figure}
    \centering
    \epsscale{1}
    \plotone{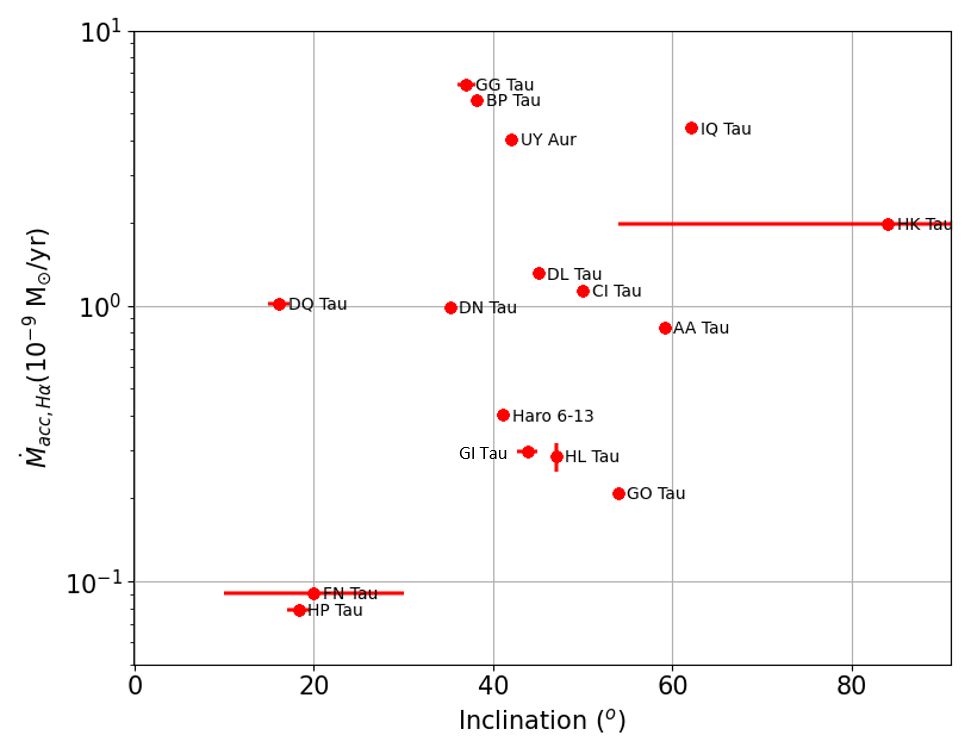}
    \caption{The relationship between the inclination and mass--accretion rates derived from the  H$\alpha$ emission luminosities. The Pearson’s correlation coefficient is about 0.34, indicating the correlation is not significant.}
    \label{fig:ha_macc_vs_inclination}
\end{figure}

The inclination angle of a CTTS may have an impact on the estimation of its mass--accretion rate \textnormal{to some degree.} 
This is because the varying geometric angle of the disk could obscure areas emitting accretion--associated emission from the line of sight, leading to uncertainties in the measurement. 
However, our analysis indicates that there is no significant correlation between the inclination angle and the log spectra--derived mass--accretion rates (see Figure~\ref{fig:ha_macc_vs_inclination}) with a Pearson's correlation coefficient of 0.34 for H$\alpha$'s measurement, and the correlation coefficients are even lower for the results from other emissions.
One possibility is that the emission regions driven by accretion are uniformly distributed on the surface of the CTTS, making the effect of disk obscuration negligible.
Another possible explanation is that the properties of the inner disk may play a more significant role than those of the outer disk. 
Several studies have reported misalignments between the inner and outer disks. 
For example, the inner disk of AA~Tau has an inclination angle of $i=75$~deg, closer to edge-on \citep[][]{2003A&A...409..169B, 2005MNRAS.358..632O}, while the outer disk has an inclination angle of $i=59$~deg \citep[][]{2017ApJ...840...23L}.
Unfortunately, information on the inclination angle of the inner disk for our sample stars, except for AA~Tau, is not available. 
Further studies using high--resolution double--peaked Keplerian line spectroscopic observations or spatially resolved observations using infrared interferometry on VLTI may be required to address this issue.

\begin{figure}
    \centering
    \epsscale{1.2}
    \gridline{\fig{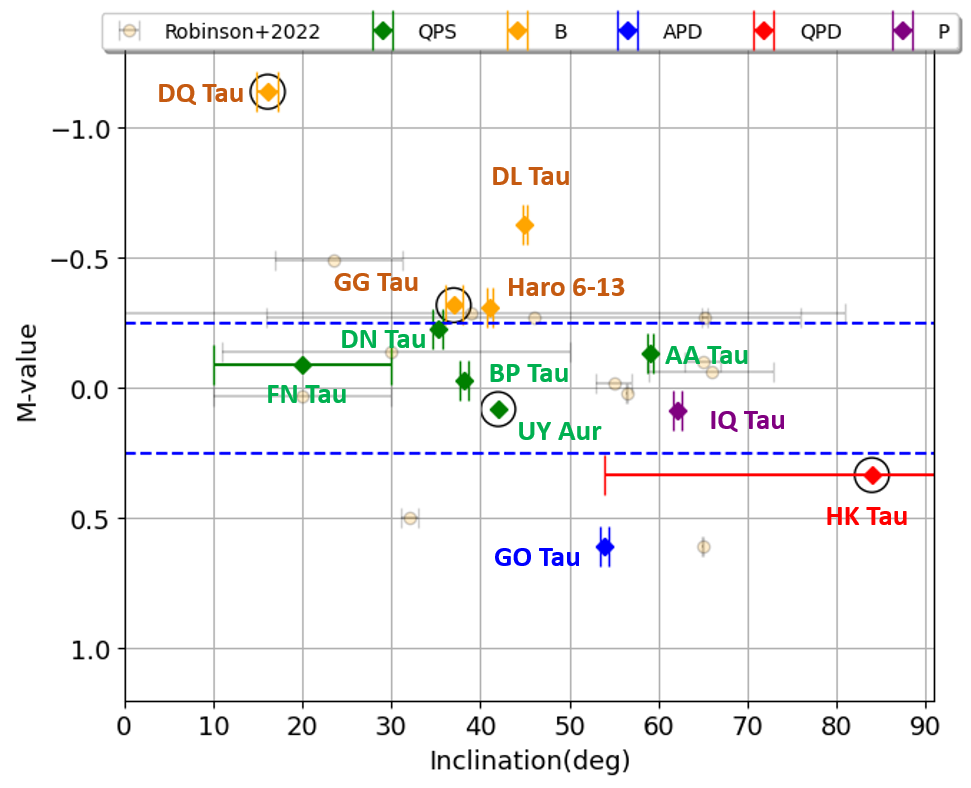}{0.65\textwidth}{}}
    \gridline{\fig{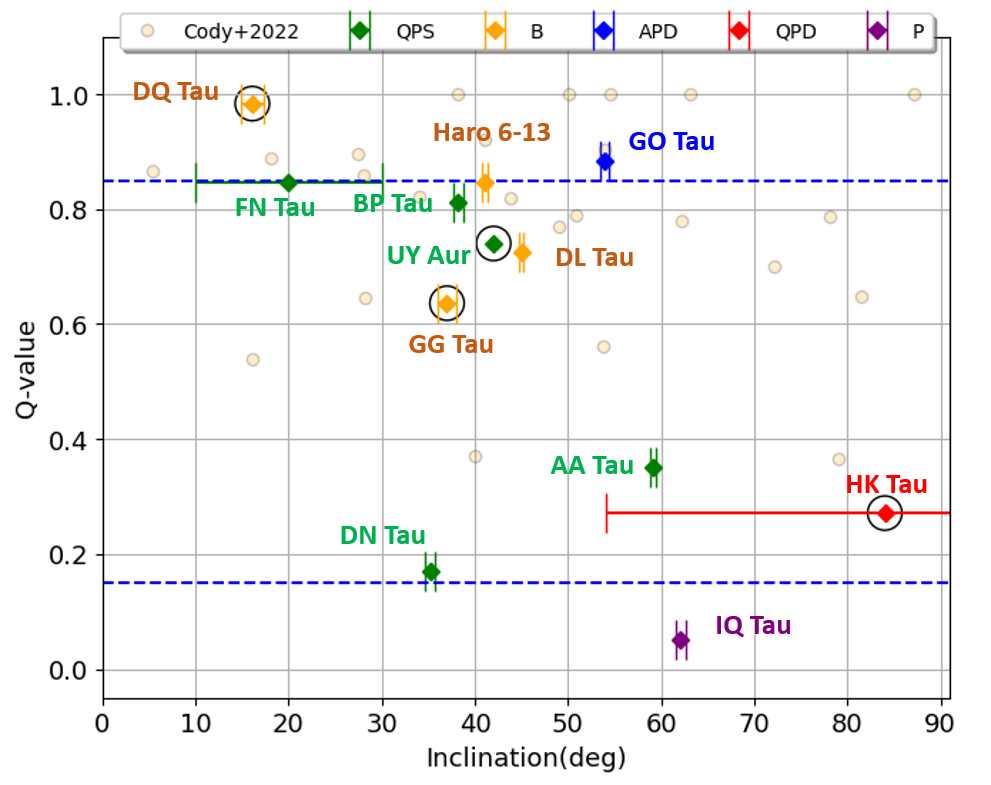}{0.65\textwidth}{}}
    \caption{The relationship between the flux asymmetry metric $M$ and the disk inclination of our CTTSs is shown in the upper panel, and the relationship between the periodicity metric $Q$ and the disk inclination is shown in the bottom panel.  Our samples are represented by rhombus symbols, with different colors indicating different variability classes. 
    \textnormal{We also include the results obtained by \cite{2022ApJ...935...54R} for $M$-vs-$inclination$ and by \cite{2022AJ....163..212C} for $Q$-vs-$inclination$ as the comparisons, shown as yellow circles.
    }
    The stars surrounded by black hollow circles in this diagram are binary or multi--star systems.
    Note that the large uncertainty of HK~Tau's inclination is because it is an unresolved binary system in LAMOST observation, which is composed of two companions, HK~Tau~A with $i=56.9^{o}$ and HK~Tau~B with $i=84^{o}$. The blue dashed lines in both panel indicate the thresholds of the bursting ($M<-0.25$), dipping ($M>0.25$), periodic ($Q<0.15$), and aperiodic ($Q>0.85$). }
    \label{fig:M+Q_vs_inclination}
\end{figure}

The type of photometric variability of CTTS is dominated by the accretion and dark spots located on the photosphere and the disk around, thus it could be geometric viewing angle dependent.
\textnormal{\cite{2018AJ....156...71C} found a positive correlation between the \emph{M} value they estimated and the inclination \cite{2017ApJ...851...85B} gave, so that dippers and QPS sources tend to have high inclinations, while bursters and stochastic sources have low inclinations.
However, due to the substantial uncertainties associated with the inclination measurements, they cautiously stated this observed tendency.}
A more recent study by \cite{2022ApJ...935...54R}, based on CTTSs with lower inclination uncertainties, found little evidence for a strong correlation between disk inclination and \emph{M} value (see Fig.12 in their paper).
\textnormal{The upper panel shown in Figure~\ref{fig:M+Q_vs_inclination} illustrates the relationship between inclination and \emph{M} value of CTTS, including our samples and those from \cite{2022ApJ...935...54R}.}
Our analysis reveals a relatively strong and positive correlation between \emph{M} value and inclination, with a Pearson's value of $0.61\pm0.03$, compared to the results given by \cite{2018AJ....156...71C} and \cite{2022ApJ...935...54R}. 

\textnormal{
On the other hand, we find an indistinct and negative correlation between the periodicity metric \emph{Q} value and inclination with a Pearson's value of $-0.54\pm0.03$ (see the bottom panel in Figure~\ref{fig:M+Q_vs_inclination}) for our targets.
This is consistent with the conclusion given by \cite{2022ApJ...935...54R} who also found the weak inverse correlation between disk inclination and periodicity metric of Taurus CTTSs.
Furthermore, the distribution of variability classes in the inclination--Q space in our study is similar to that of Taurus members studied by \cite{2022AJ....163..212C}, as shown in Figure~7 of their paper.
Nevertheless, it is possible that such distribution and the correlation between \emph{M} value and inclination, as well as the correlation between periodicity metric \emph{Q} value and inclination, could vary over time due to the changing variability classes of CTTSs.
}
\section{Summary}
\label{sec:section5}
We present a comprehensive study of \textnormal{16} low--mass classical T Tauri stars in TMC for their mass--accretion and luminosity variability properties by using LAMOST low--resolution spectra and TESS photometric light curves.

\textnormal{From the LAMOST spectra, the mass--accretion rates of 16 CTTSs have been estimated using H$\alpha$, H$\beta$, Ca$\text{II}$, and He$\text{I}$ emissions as indicators.
The average mass--accretion rate of all our samples based on H$\alpha$ lines is found to be $1.76\times10^{-9} M_{\odot}yr^{-1}$.
Not all targets in the sample have measurable equivalent widths for the other emission lines, which means that not all corresponding mass--accretion rates could be estimated. 
HL~Tau has the largest dispersion of the spectral--derived mass-accretion rates with a coefficient of variation (CV) value of $1.15$, which is the ratio of the standard deviation to the mean. 
The reason that causes this huge dispersion remains unclear.
On the other hand, the lowest dispersion CV value is found in FN~Tau, which is $0.08$.
}

\textnormal{We also compute the time--series mass--accretion rates based on the TESS light curve, ZTF, and ASAS--SN survey for the stars DL~Tau and Haro~6--13, which are identified to be the bursters according to their brightness variability asymmetry metric \emph{M} values of $-0.63$  and $-0.31$, respectively \citep{2014AJ....147...82C}.
In total over about 50 days, the average accretion rates of DL~Tau and Haro~6--13 are $2.5\times10^{-9} M_{\odot}yr^{-1}$ and $1.4\times10^{-11} M_{\odot}yr^{-1}$, respectively.
The rate of DL~Tau is consistent with the mean value of the mass--accretion derived from LAMOST.
However, for Haro~6--13, this rate is generally lower compared to the LAMOST estimate
We further investigate whether the mass--accretion mechanisms of these two stars are dominated by steady or burst accretion.
By assuming the average rates obtained from LAMOST emission measurement as the steady rates, the mass accretions in DL~Tau and Haro~6--13 are dominated by the steady process.}

\textnormal{We have detected 13 flares in total in BP~Tau, DN~Tau, and FN~Tau.
The amplitude, duration, timing at peak in BTJD, and released energy of detected flares are determined, with the most energetic flare observed in FN~Tau with the energy of $6.3\times10^{35}$ erg.
The power--law flare frequency distribution with a slope of $\beta=-0.68\pm0.20$ indicates that the flare activity of the CTTSs is dominated by strong flare with high energy.
Our analysis of the flare frequency distribution also reveals that the stars in our sample display extremely high levels of flare activity with the capability of producing flares with energy greater than $E=1\times10^{34}$~erg up to 12 times per year, which is a significant deviation from solar--type stars and young M dwarfs.
}

\textnormal{
Ten stars in our sample were previously observed and classified for variability by \cite{2022AJ....163..212C} and \cite{2022ApJ...935...54R}. 
Comparing the variability classes from \cite{2022AJ....163..212C} based on K2 Campaign 13 data from about four years ago, we found that most of the stars (6 out of 7) showed changes in their variability, except for DQ~Tau.
In comparison with the results from \cite{2022ApJ...935...54R}, BP~Tau and FN~Tau's variability classes remained the same, while UY~Aur changed from "B" to "QPS" in 600 days but still exhibited a burst--like event in its TESS Section 43 light curve. 
These findings suggest that the long-term timescale of variation in CTTSs could be between 1.6 to 4 years or longer and require further investigation to determine if changes occur cyclically or randomly.
}

The CTTSs' projected wind velocities in terms of the blue edge/peaks of [OI], [SII], and [CaII] forbidden emissions detected in LAMOST are found to be correlated with the inclination, that is stars with low inclination tend to have fast projected wind velocity.
On the other hand, no tendency has been found between inclination and mass--accretion rates derived from the spectral lines' luminosities.
\textnormal{One possibility is that the regions of emission driven by accretion are distributed uniformly across the stellar surface, so the disk obscuration doesn't have much of an impact.}
Alternatively, it's possible that the properties of the inner disk are more important than those of the outer disk. 
There is evidence to suggest that the inner and outer disks of CTTS can become misaligned due to, e.g., magnetic warping.
Unfortunately, in our sample, only AA~Tau has reported inner disk inclination angle.
High--resolution double--peaked Keplerian line spectroscopic observations or spatially resolved observations using infrared interferometry on VLTI may be required for further studies of this topic.

\textnormal{The brightness variabilities of CTTSs have been also considered to be influenced by the inclinations to some degree.}
Our analysis reveals a relatively significant and positive correlation between \emph{M} value and inclination, with a Pearson's value of $0.61\pm0.03$, compared to the results given by \cite{2018AJ....156...71C} and \cite{2022ApJ...935...54R}. 
We observed a weak negative correlation between the periodicity metric \emph{Q} value and inclination for our targets, with a Pearson's value of $-0.54\pm0.03$.
This finding supports the conclusion reached by \cite{2022ApJ...935...54R}, who also reported a weak inverse correlation between disk inclination and the periodicity metric of Taurus CTTSs. 
Additionally, we found that the distribution of variability classes in the inclination--Q space is similar to that of Taurus members studied by \cite{2022AJ....163..212C}. 
However, since the variability classes of CTTSs can change over time, it is possible that the distribution and correlations between the \emph{M} value and inclination, and the periodicity metric \emph{Q} value and inclination, could vary as well. 
Thus, further research is required to fully understand the long-term evolution of these correlations.

\begin{acknowledgments}

We thank the anonymous referee for giving constructive comments.
This research is supported in part by grant No.~107-2119-M-008-012 of MOST, Taiwan.
This paper includes data collected by the TESS, ZTF, ASAS-SN, Pan-STARRS1, and LAMOST. \

Funding for the TESS mission is provided by the NASA's Science Mission Directorate.
All the TESS data used in this paper can be found in MAST: \dataset[10.17909/t9-nmc8-f686]{http://dx.doi.org/10.17909/t9-nmc8-f686}.

ZTF is supported by National Science Foundation grant AST-1440341 and a collaboration including Caltech, IPAC, the Weizmann Institute for Science, the Oskar Klein Center at Stockholm University, the University of Maryland, the University of Washington, Deutsches ElektronenSynchrotron and Humboldt University, Los Alamos National Laboratories, the TANGO Consortium of Taiwan, the University of Wisconsin at Milwaukee, and Lawrence Berkeley National Laboratories. Operations are conducted by COO, IPAC, and UW \citep{ZTF2022}.

ASAS-SN is supported by the Gordon and Betty Moore Foundation through grant GBMF5490 to the Ohio State University and NSF grant AST-1515927. Development of ASAS-SN has been supported by NSF grant AST-0908816, the Mt. Cuba Astronomical Foundation, the Center for Cosmology and Astro- Particle Physics at the Ohio State University, the Chinese Academy of Sciences South America Center for Astronomy (CASSACA), the Villum Foundation, and George Skestos.

The Pan-STARRS1 Surveys (PS1) and the PS1 public science archive have been made possible through contributions by the Institute for Astronomy, the University of Hawaii, the Pan-STARRS Project Office, the Max-Planck Society and its participating institutes, the Max Planck Institute for Astronomy, Heidelberg and the Max Planck Institute for Extraterrestrial Physics, Garching, The Johns Hopkins University, Durham University, the University of Edinburgh, the Queen's University Belfast, the Harvard-Smithsonian Center for Astrophysics, the Las Cumbres Observatory Global Telescope Network Incorporated, the National Central University of Taiwan, the Space Telescope Science Institute, the National Aeronautics and Space Administration under Grant No. NNX08AR22G issued through the Planetary Science Division of the NASA Science Mission Directorate, the National Science Foundation Grant No. AST–1238877, the University of Maryland, Eotvos Lorand University (ELTE), the Los Alamos National Laboratory, and the Gordon and Betty Moore Foundation.

The Guoshoujing Telescope (the Large Sky Area Multi-Object Fiber Spectroscopic Telescope, LAMOST) is a National Major Scientific Project built and operated by the Chinese Academy of Sciences.

This research made use of Lightkurve, a Python package for Kepler and TESS data analysis \citep{2018ascl.soft12013L}.

\end{acknowledgments}
\software{Astropy \citep{2013A&A...558A..33A, 2018AJ....156..123A, 2022ApJ...935..167A}, Matplotlib \citep{hunter2007matplotlib}, NumPy \citep{oliphant2006guide}, Pandas \citep{mckinney2010data}, Scipy \citep{virtanen2020scipy}}

\clearpage

\begin{deluxetable}{c c c c c c c c c c c c c c c}
\tabletypesize{\normalsize}
\tablecaption{\\16 CTTSs in TMC observed by LAMOST and TESS.}

\tablecolumns{19}
\tablewidth{0.1pt}
\tablehead{
\colhead{} & 
\colhead{Name} & 
\colhead{LAMOST Obser. Date} &
\colhead{TIC} 
\\
\colhead{} & 
\colhead{(1)} & 
\colhead{(2)} &
\colhead{(3)} 
}
\startdata
     1 &    AA Tau &      2019-01-29 & 268510757 \\
     2 &    BP Tau &      2014-01-25 &  58287935 \\
     3 &    CI Tau &      2016-02-07 &  61230756 \\
     4 &    DL Tau &      2012-12-23 & 268444139 \\
     5 &    DN Tau &      2012-12-23 & 268511247 \\
     6 &    DQ Tau &      2020-01-26 & 436614017 \\
     7 &    FN Tau &      2014-01-25 &  56625009 \\
     8 &    GG Tau &      2016-02-05 & 245830955 \\
     9 &    GO Tau &      2013-11-29 & 125914458 \\
    10 & Haro 6-13 &      2012-12-23 & 268324608 \\
    11 &    HK Tau &      2012-11-07 & 268324063 \\
    12 &    HL Tau &      2013-02-09 & 353752575 \\
    13 &    IQ Tau &      2012-12-23 & 268219495 \\
    14 &    UY Aur &      2019-12-30 &  96205005 \\
    15 &    GI Tau &      2019-01-29 & 268444803 \\
    16 &    HP Tau &      2012-12-23 & 118521708 \\
\enddata

\tablenotetext{}{}

\label{tab:tab1}

\end{deluxetable}

\begin{deluxetable}{c c c c c c c c c c c c c c c}
\tabletypesize{\normalsize}
\tablecaption{\\Stellar properties of 16 CTTSs in this study}

\tablecolumns{19}
\tablewidth{0.1pt}
\tablehead{
\colhead{Name} & 
\colhead{Radius} &
\colhead{Mass} &
\colhead{Spec. type} &
\colhead{Distance} & 
\colhead{Inclination} 
\\
\colhead{} & 
\colhead{ (R$_{\odot}$)  } & 
\colhead{ (M$_{\odot}$)} & 
\colhead{} & 
\colhead{(pc)} & 
\colhead{(deg)} & 
\\
\colhead{(1)} & 
\colhead{(2)} & 
\colhead{(3)} & 
\colhead{(4)} & 
\colhead{(5)} & 
\colhead{(6)} 
}
\startdata
   AA Tau & 1.72 &  0.55 &   M0.6 &       134.671 &              59.1$\pm$0.3 \\
   BP Tau & 1.30 &  0.63 &   M0.5 &       127.398 &              38.2$\pm$0.5 \\
   CI Tau & 1.53 &  0.90 &   K5.5 &       160.318 &              50.0$\pm$0.3 \\
   DL Tau & 1.20 &  0.94 &   K5.5 &       159.936 &              45.0$\pm$0.2 \\
   DN Tau & 2.05 &  0.55 &   M0.3 &       128.601 &              35.2$\pm$0.5 \\
   DQ Tau & 1.77 &  0.54 &   M0.6 &       195.381 &              16.1$\pm$1.2 \\
   FN Tau & 2.30 &  0.29 &   M3.5 &       129.894 &              20.0$\pm$10.0 \\
   GG Tau & 2.39 &  0.63 &   K7.5 &       116.395 &              37.0$\pm$1.0 \\
   GO Tau & 1.25 &  0.41 &   M2.3 &       142.383 &              53.9$\pm$0.5 \\
Haro 6-13 & 1.83 &  0.84 &   K5.5 &       128.555 &              41.1$\pm$0.3 \\
   HK Tau* & 1.36 &  0.49 &   M1.5 &      140.000 &              56.9(HK~Tau~A), 84.0 (HK~Tau~B) \\
   HL Tau & 7.00 &  1.70 &     K3 &       140.000 &              47.0$\pm$0.4 \\
   IQ Tau & 1.55 &  0.51 &   M1.1 &       131.511 &              62.1$\pm$0.5 \\
   UY Aur & 1.90 &  0.70 &   K7.0 &       152.279 &              42.0 \\
   GI Tau & 1.72 &  0.58 &   M0.4 &       129.436 &              43.8$\pm$1.1 \\
   HP Tau & 1.81 &  1.06 &   K4.0 &       171.248 &              18.3$\pm$1.3 \\
\enddata

\tablenotetext{}{\textnormal{Note.} The masses and radii of most CTTSs are provided in \cite{2014ApJ...786...97H}, with the exception of HL~Tau.
The radius (R=7~R$_{\odot}$) and mass (M=1.7M$_{\odot}$) of HL~Tau are given by \cite{Liu et al.(2017)} and \cite{2016ApJ...816...25P}, respectively. The spectral types are obtained from LAMOST spectral pipeline.
The distances are from the parallax measurements reported in \cite{2020yCat.1350....0G}. 
The references of the inclinations: BP~Tau, CI~Tau, DL~Tau, DN~Tau, DQ~Tau, GO~Tau, Haro~6--13, IQ~Tau, GI~Tau, and HP~Tau \citep{2019ApJ...882...49L}; HL~Tau \citep{2015ApJ...808L...3A}; AA~Tau \citep{2017ApJ...840...23L}; GG~Tau \citep{Guilloteau1999}; HK~Tau~A \citep{Manara et al.(2019)} and HK~Tau~B \citep{Villenave2020}; UY~Aur \citep{1998ApJ...499..883C}. Note that the stellar parameters, mass and radius, of HK~Tau shown here belongs to HK~Tau~A.}

\label{tab:tab2}

\end{deluxetable}

\begin{deluxetable}{c c c c c c c c c c c c c c c}
\tabletypesize{\normalsize}
\tablecaption{\\Photometric data of our samples taken by ASAS--SN and Pan–-STARRS surveys within 20 days of LAMOST observation.}

\tablecolumns{19}
\tablewidth{0.1pt}
\tablehead{
\colhead{} & 
\colhead{Name} & 
\colhead{LAMOST OBT} &
\colhead{ASAS--SN OBT} &
\colhead{ASAS--SN $g$} &
\colhead{ASAS--SN $V$} & 
\colhead{Pan--Starrs OBT} & 
\colhead{Pan--Starrs $g$} &
\colhead{Pan--Starrs $i$} 
\\
\colhead{} & 
\colhead{} & 
\colhead{(MJD)} &
\colhead{(MJD)  } & 
\colhead{ (mJy)} & 
\colhead{(mJy)} & 
\colhead{(MJD)} & 
\colhead{(mJy)} & 
\colhead{(mJy)} 
\\
\colhead{} & 
\colhead{(1)} & 
\colhead{(2)} &
\colhead{(3)} &
\colhead{(4)} & 
\colhead{(5)} & 
\colhead{(6)} & 
\colhead{(7)} &
\colhead{(8)} 
}
\startdata
     1 &    AA Tau &          58512.516 &    58512.037 &          0.88$\pm$0.06 &             ...&           ...&              ...&              ...\\
     2 &    BP Tau &          56682.458 &    56700.294 &             ...&          48.58$\pm$0.18 &           ...&              ...&              ...\\
     3 &    CI Tau &          56682.458 &    57424.345 &             ...&          16.49$\pm$0.14 &           ...&              ...&              ...\\
     4 &    DL Tau &          56284.592 &          ...&             ...&             ...&    53600.366 &            10.37$\pm$0.01 &              ...\\
     5 &    DN Tau &          56284.609 &          ...&             ...&             ...&    56333.298 &            26.09$\pm$0.02 &              ...\\
     6 &    GO Tau &          56625.627 &          ...&             ...&             ...&    56605.593 &              ...&            47.64$\pm$0.06 \\
    7 & Haro 6-13 &          56284.597 &          ...&             ...&             ...&    56300.366 &             1.25$\pm$0.01 &              ...\\
    8 &    HK Tau &          56238.778 &          ...&             ...&             ...&    56236.566 &              ...&            52.60$\pm$0.06 \\
    9 &    HL Tau &          56332.471 &          ...&             ...&             ...&    56333.325 &             1.57$\pm$0.01 &              ...\\
    10 &    IQ Tau &          56284.607 &          ...&             ...&             ...&    56300.365 &            49.00$\pm$0.04 &              ...\\
\enddata

\tablenotetext{}{}

\label{tab:tab_asas-sn+panstarr}

\end{deluxetable}

\begin{deluxetable}{c c c c c c c c c c c c c c c}
\tabletypesize{\normalsize}
\tablecaption{The photometric variability classes of the CTTSs observed by TESS in this study and the literature (if any).}

\tablecolumns{19}
\tablewidth{0.1pt}
\tablehead{
\colhead{Name} & 
\colhead{\emph{M*}} &
\colhead{\emph{M$^{C}$}} &
\colhead{\emph{M$^{R}$}} &
\colhead{\emph{Q*}} & 
\colhead{\emph{Q$^{C}$}} & 
\colhead{\emph{Q$^{R}$}}&
\colhead{Variability*} &
\colhead{Variability$^{C}$} &
\colhead{Variability$^{R}$} &
\colhead{Period$^{*}$} & 
\colhead{Period$^{P}$} &
\colhead{Period$^{Re}$} 
\\
\colhead{(1)} & 
\colhead{(2)} &
\colhead{(3)} &
\colhead{(4)} & 
\colhead{(5)} & 
\colhead{(6)} & 
\colhead{(7)} &
\colhead{(8)} &
\colhead{(9)} & 
\colhead{(10)} &
\colhead{(11)} &
\colhead{(12)} &
\colhead{(13)}
}
\startdata
   AA Tau &    -0.13 &         -0.30 &               ... &         0.35 &          1.00 &               ... &        QPS &          S &  ... & 8.3  & 8.2  & ...\\
   BP Tau &    -0.03 &           ... &             -0.03 &         0.81 &           ... &              0.83 &      QPS &        ... &            QPS & 7.6  & 7.6  & ...\\
   DL Tau &    -0.63 &           ... &               ... &         0.72 &           ... &               ... &        B &        ... &            ... & 10.4  & 9.35  & ...\\
   DN Tau &    -0.23 &           ... &               ... &         0.17 &           ... &               ... &      QPS &        ... &            ... & 6.18  & 6.1  & ...\\
   DQ Tau &    -1.14 &         -0.87 &               ... &         0.98 &          0.54 &               ... &        B &          B &            ... & 3.03  & ...  & 3.03\\
   FN Tau &    -0.09 &           ... &              0.03 &         0.85 &           ... &              0.60 &      QPS &        ... &            QPS & 7.33  & ...  & 8.77 \\
   GG Tau &    -0.32 &           ... &               ... &         0.64 &           ... &               ... &        B &        ... &            ... & 10.3  & 10.3  & ...\\
   GO Tau &     0.61 &          0.25 &               ... &         0.88 &          0.90 &               ... &      APD &          U &            ... & 2.51  & ...  & ...\\
Haro 6-13 &    -0.31 &          0.10 &               ... &         0.85 &          0.92 &               ... &        B &          S &            ... & 3.27  & ...  & 3.27 \\
   HK Tau &     0.33 &          0.11 &               ... &         0.27 &          0.72 &               ... &      QPD &        QPS &            ... & 3.31  & 3.31  & ...\\
   IQ Tau &     0.08 &         -0.13 &               ... &         0.05 &          0.78 &               ... &      P &        QPS &            ... & 12.92  & ...  & 6.67\\
   UY Aur &     0.08 &           ... &             -0.49 &         0.74 &           ... &              0.74 &      QPS &        ... &              B & 3.04  & ...  & ...\\
\enddata

\tablenotetext{}{\textnormal{Note.} $*$ represents the measurements given in this study. $^{C}$: \cite{2022AJ....163..212C}.  $^{R}$: \cite{2022ApJ...935...54R}.$^{P}$: \cite{2010PASP..122..753P}. $^{Re}$: \cite{2020AJ....159..273R}. The unit of the period is in days.}

\label{tab:M+Q_value}

\end{deluxetable}


\movetabledown=6.5cm
\begin{rotatetable}
\begin{deluxetable*}{cD@{$\pm$}D D@{$\pm$}D D@{$\pm$}D D@{$\pm$}D D@{$\pm$}D D@{$\pm$}D D@{$\pm$}D D@{$\pm$}D D@{$\pm$}D D@{$\pm$}D}
\tabletypesize{\normalsize}
\tablecaption{\\Equivalent widths of all mass--accretion indicator emissions of 16 CTTSs in LAMOST spectra in this study.}

\tablehead{
\colhead{Name} & 
\multicolumn4c{$EW_{H\alpha}$} & 
\multicolumn4c{$EW_{H\beta}$} &
\multicolumn4c{$EW_{Ca\text{II} 8498}$} &
\multicolumn4c{$EW_{Ca\text{II} 8542}$} &
\multicolumn4c{$EW_{Ca\text{II} 8662}$} &
\multicolumn4c{$EW_{He\text{I} 4771}$} &
\multicolumn4c{$EW_{He\text{I} 5875}$} &
\multicolumn4c{$EW_{He\text{I} 6678}$} &
\multicolumn4c{$EW_{He\text{I} 7065}$} 
\\
\colhead{(1)} & 
\multicolumn4c{(2)} & 
\multicolumn4c{(3)} & 
\multicolumn4c{(4)} & 
\multicolumn4c{(5)} & 
\multicolumn4c{(6)} & 
\multicolumn4c{(7)} &
\multicolumn4c{(8)} &
\multicolumn4c{(9)} &
\multicolumn4c{(10)}
}

\decimals
\startdata
   AA Tau & 66.45 &    1.16 & 18.96 &    0.33 &         ... &           ... &         ... &           ... &         ... &           ... &         2.14 &           0.06 &         3.00 &           0.07 &         0.66 &           0.02 &          ... &            ... \\
   BP Tau & 98.29 &    0.06 & 23.15 &    0.01 &         ... &           ... &         ... &           ... &         ... &           ... &         2.98 &           0.05 &          ... &            ... &          ... &            ... &          ... &            ... \\
   CI Tau & 44.39 &    2.07 & 13.18 &    0.61 &       11.74 &          0.10 &       10.18 &          0.06 &        8.68 &          0.01 &         0.96 &           0.02 &         2.17 &           0.06 &         0.73 &           0.04 &          ... &            ... \\
   DL Tau & 83.17 &    0.78 & 28.41 &    0.27 &       40.52 &          2.43 &       44.29 &          0.00 &       37.03 &          0.60 &         2.63 &           0.08 &         5.45 &           0.03 &         1.92 &           0.05 &         1.78 &           0.03 \\
   DN Tau & 17.06 &    0.76 &  9.70 &    0.43 &         ... &           ... &         ... &           ... &         ... &           ... &          ... &            ... &         0.85 &           0.05 &          ... &            ... &          ... &            ... \\
   DQ Tau & 81.36 &    1.31 & 20.44 &    0.33 &         ... &           ... &         ... &           ... &         ... &           ... &         1.89 &           0.15 &         1.73 &           0.11 &          ... &            ... &          ... &            ... \\
   FN Tau & 22.36 &    0.53 & 11.57 &    0.27 &         ... &           ... &         ... &           ... &         ... &           ... &         0.82 &           0.00 &         1.34 &           0.03 &          ... &            ... &          ... &            ... \\
   GG Tau & 59.92 &    1.35 & 26.44 &    0.60 &        2.12 &          0.05 &        2.08 &          0.02 &        1.91 &          0.06 &          ... &            ... &         1.55 &           0.00 &          ... &            ... &          ... &            ... \\
   GO Tau & 36.64 &    2.01 & 20.68 &    1.13 &         ... &           ... &         ... &           ... &         ... &           ... &         2.74 &           0.06 &         1.73 &           0.13 &          ... &            ... &          ... &            ... \\
Haro 6-13 & 73.51 &    1.74 &   ... &     ... &        4.97 &          0.00 &        4.59 &          0.00 &        4.16 &          0.00 &          ... &            ... &          ... &            ... &          ... &            ... &          ... &            ... \\
   HK Tau & 57.27 &    0.97 & 25.12 &    0.42 &         ... &           ... &         ... &           ... &         ... &           ... &          ... &            ... &         1.75 &           0.04 &          ... &            ... &          ... &            ... \\
   HL Tau & 59.73 &   10.95 & 10.67 &    1.96 &       27.57 &          0.60 &       28.72 &          0.00 &       24.13 &          0.29 &          ... &            ... &          ... &            ... &          ... &            ... &          ... &            ... \\
   IQ Tau & 24.67 &    0.84 & 13.81 &    0.47 &         ... &           ... &         ... &           ... &         ... &           ... &         2.37 &           0.16 &         2.13 &           0.04 &         0.66 &           0.02 &          ... &            ... \\
   UY Aur & 46.27 &    0.97 & 16.32 &    0.34 &        3.36 &          0.07 &        2.74 &          0.05 &        2.74 &          0.00 &         1.43 &           0.06 &         1.42 &           0.04 &          ... &            ... &          ... &            ... \\
   GI Tau & 25.33 &    1.60 & 14.41 &    0.91 &        1.34 &          0.00 &        1.04 &          0.07 &        1.51 &          0.00 &         2.89 &           0.06 &         3.07 &           0.03 &         1.51 &           0.02 &         0.91 &           0.00 \\
   HP Tau & 12.44 &    0.59 &  2.13 &    0.10 &         ... &           ... &         ... &           ... &         ... &           ... &          ... &            ... &          ... &            ... &          ... &            ... &          ... &            ... \\
\enddata

\tablenotetext{}{\textnormal{Note.} Note that we use positive equivalent width to indicate emission. The unit is $\AA$}
\label{tab:spectra-EW}
\end{deluxetable*}
\end{rotatetable}

\movetabledown=6.5cm
\begin{rotatetable}
\begin{deluxetable*}{cD@{$\pm$}D D@{$\pm$}D D@{$\pm$}D D@{$\pm$}D D@{$\pm$}D D@{$\pm$}D D@{$\pm$}D D@{$\pm$}D D@{$\pm$}D D@{$\pm$}D}
\tabletypesize{\normalsize}
\tablecaption{\\Mass--accretion rates of 16 CTTSs derived from LAMOST spectra in this study.}

\tablehead{
\colhead{Name} & 
\multicolumn4c{$\dot{M}_{H\alpha}$} & 
\multicolumn4c{$\dot{M}_{H\beta}$} &
\multicolumn4c{$\dot{M}_{Ca\text{II} 8498}$} &
\multicolumn4c{$\dot{M}_{Ca\text{II} 8542}$} &
\multicolumn4c{$\dot{M}_{Ca\text{II} 8662}$} &
\multicolumn4c{$\dot{M}_{He\text{I} 4771}$} &
\multicolumn4c{$\dot{M}_{He\text{I} 5875}$} &
\multicolumn4c{$\dot{M}_{He\text{I} 6678}$} &
\multicolumn4c{$\dot{M}_{He\text{I} 7065}$} 
\\
\colhead{(1)} & 
\multicolumn4c{(2)} & 
\multicolumn4c{(3)} & 
\multicolumn4c{(4)} & 
\multicolumn4c{(5)} & 
\multicolumn4c{(6)} & 
\multicolumn4c{(7)} &
\multicolumn4c{(8)} &
\multicolumn4c{(9)} &
\multicolumn4c{(10)}
}

\decimals
\startdata
   AA Tau &             0.835 &               0.010 &             1.399 &               0.016 &                   ... &                     ... &                   ... &                     ... &                   ... &                     ... &                    1.713 &                      0.028 &                    1.625 &                      0.026 &                    0.845 &                      0.021 &                      ... &                        ... \\
   BP Tau &             5.591 &               0.002 &             5.831 &               0.002 &                   ... &                     ... &                   ... &                     ... &                   ... &                     ... &                    8.488 &                      0.092 &                      ... &                        ... &                      ... &                        ... &                      ... &                        ... \\
   CI Tau &             1.133 &               0.035 &             0.889 &               0.027 &                 6.902 &                   0.033 &                 4.994 &                   0.016 &                 4.567 &                   0.004 &                    0.745 &                      0.009 &                    2.154 &                      0.037 &                    2.588 &                      0.095 &                      ... &                        ... \\
   DL Tau &             1.314 &               0.008 &             2.090 &               0.013 &                 9.859 &                   0.353 &                 5.212 &                   0.000 &                 6.757 &                   0.061 &                    2.714 &                      0.050 &                    3.541 &                      0.015 &                    4.527 &                      0.080 &                    5.378 &                      0.055 \\
   DN Tau &             0.993 &               0.029 &             2.745 &               0.081 &                   ... &                     ... &                   ... &                     ... &                   ... &                     ... &                      ... &                        ... &                    2.252 &                      0.090 &                      ... &                        ... &                      ... &                        ... \\
   DQ Tau &             1.022 &               0.011 &             0.831 &               0.009 &                   ... &                     ... &                   ... &                     ... &                   ... &                     ... &                    0.860 &                      0.043 &                    0.706 &                      0.030 &                      ... &                        ... &                      ... &                        ... \\
   FN Tau &             0.091 &               0.001 &             0.087 &               0.001 &                   ... &                     ... &                   ... &                     ... &                   ... &                     ... &                    0.099 &                      0.000 &                    0.107 &                      0.001 &                      ... &                        ... &                      ... &                        ... \\
   GG Tau &             6.349 &               0.095 &             9.110 &               0.136 &                 3.579 &                   0.052 &                 2.901 &                   0.017 &                 2.961 &                   0.051 &                      ... &                        ... &                    5.937 &                      0.003 &                      ... &                        ... &                      ... &                        ... \\
   GO Tau &             0.209 &               0.007 &             0.323 &               0.012 &                   ... &                     ... &                   ... &                     ... &                   ... &                     ... &                    0.592 &                      0.008 &                    0.289 &                      0.014 &                      ... &                        ... &                      ... &                        ... \\
Haro 6-13 &             0.402 &               0.006 &               ... &                 ... &                 2.424 &                   0.000 &                 1.353 &                   0.000 &                 2.083 &                   0.000 &                      ... &                        ... &                      ... &                        ... &                      ... &                        ... &                      ... &                        ... \\
   HK Tau &             1.983 &               0.022 &             2.517 &               0.028 &                   ... &                     ... &                   ... &                     ... &                   ... &                     ... &                      ... &                        ... &                    1.994 &                      0.033 &                      ... &                        ... &                      ... &                        ... \\
   HL Tau &             0.285 &               0.034 &             0.185 &               0.022 &                 3.538 &                   0.045 &                12.163 &                   0.000 &                 2.959 &                   0.019 &                      ... &                        ... &                      ... &                        ... &                      ... &                        ... &                      ... &                        ... \\
   IQ Tau &             4.450 &               0.100 &             7.774 &               0.175 &                   ... &                     ... &                   ... &                     ... &                   ... &                     ... &                   13.949 &                      0.573 &                   15.126 &                      0.174 &                   18.257 &                      0.421 &                      ... &                        ... \\
   UY Aur &             4.017 &               0.055 &             9.112 &               0.126 &                 3.915 &                   0.051 &                 2.616 &                   0.025 &                 2.842 &                   0.000 &                    9.620 &                      0.263 &                    5.433 &                      0.099 &                      ... &                        ... &                      ... &                        ... \\
   GI Tau &             0.296 &               0.012 &             1.072 &               0.045 &                 0.369 &                   0.000 &                 0.249 &                   0.009 &                 0.425 &                   0.000 &                    3.295 &                      0.043 &                    1.735 &                      0.012 &                    2.604 &                      0.022 &                    2.021 &                      0.004 \\
   HP Tau &             0.079 &               0.002 &             0.024 &               0.001 &                   ... &                     ... &                   ... &                     ... &                   ... &                     ... &                      ... &                        ... &                      ... &                        ... &                      ... &                        ... &                      ... &                        ... \\
\enddata

\tablenotetext{}{\textnormal{Note.} The unit of all measurement shown here is $\times10^{-9}$~M$_{\odot}$/yr}
\label{tab:spectra-driven-Macc}
\end{deluxetable*}
\end{rotatetable}


\begin{deluxetable}{c c c c c c c c c c c c c c c}
\tabletypesize{\normalsize}
\tablecaption{\\ 13 flares detected in the TESS light curves of BP~Tau, DN~Tau, and FN~Tau.}

\tablecolumns{19}
\tablewidth{0.1pt}
\tablehead{
\colhead{Source} &
\colhead{Amplitude} & 
\colhead{$E_{f,tess}$}  &
\colhead{$t_{start}$} & 
\colhead{$t_{peak}$} & 
\colhead{$t_{end}$} & 
\colhead{Duration} &
\\
\colhead{} & 
\colhead{} & 
\colhead{Log(erg)} & 
\colhead{(BT-2457000)} & 
\colhead{(BT-2457000)} & 
\colhead{(BT-2457000)} & 
\colhead{(days)} 
\\
\colhead{} & 
\colhead{(1)} & 
\colhead{(2)} &
\colhead{(3)} &
\colhead{(4)} & 
\colhead{(5)} & 
\colhead{(6)} & 
}
\startdata
BP Tau &      0.042 &    34.46 &               2505.0438 &              2505.0549 &             2505.0980 &        0.053 \\
BP Tau &      0.067 &    34.34 &               2515.6458 &              2515.6486 &             2515.6833 &        0.036 \\
BP Tau &      0.221 &    35.03 &               2521.2405 &              2521.2433 &             2521.2947 &        0.053 \\
DN Tau &      0.017 &    34.54 &               2497.5972 &              2497.6097 &             2497.6514 &        0.053 \\
DN Tau &      0.104 &    35.66 &               2505.9062 &              2505.9201 &             2506.1562 &        0.249 \\
DN Tau &      0.035 &    34.95 &               2514.0178 &              2514.0553 &             2514.1025 &        0.083 \\
DN Tau &      0.019 &    34.91 &               2515.5332 &              2515.5582 &             2515.6235 &        0.089 \\
DN Tau &      0.017 &    34.39 &               2515.7526 &              2515.7568 &             2515.7832 &        0.029 \\
DN Tau &      0.046 &    35.23 &               2517.1444 &              2517.1597 &             2517.3361 &        0.190 \\
FN Tau &      0.020 &    35.12 &               2497.0780 &              2497.2433 &             2497.3128 &        0.233 \\
FN Tau &      0.087 &    35.80 &               2497.7878 &              2497.8489 &             2498.0934 &        0.304 \\
FN Tau &      0.017 &    35.03 &               2516.6918 &              2516.7668 &             2516.8987 &        0.206 \\
FN Tau &      0.023 &    35.21 &               2516.8987 &              2517.0015 &             2517.1085 &        0.208 \\
\enddata

\tablenotetext{}{\textnormal{Note:} \textnormal{The columns are: (1) flare peak amplitude, (2) flare energy in the TESS band, (3) start time of flare, (4) time of flare peak, (5) end time of flare, and (6) duration of flare in the unit of days.}
}
\label{tab:flares}
\end{deluxetable}

\begin{deluxetable*}{cD@{$\pm$}D D@{$\pm$}D D@{$\pm$}D D@{$\pm$}D D@{$\pm$}D D@{$\pm$}D D@{$\pm$}D D@{$\pm$}D D@{$\pm$}D D@{$\pm$}D}
\tabletypesize{\normalsize}
\tablecaption{\\The projected wind velocities of CTTSs observed by LAMOST.}
\tablehead{
\colhead{Name} & 
\multicolumn{4}{c}{[OI]~6300 $v_{wind}$} & 
\multicolumn{4}{c}{[SII]~6716 $v_{wind}$} & 
\multicolumn{4}{c}{[SII]~6731 $v_{wind}$} & 
\multicolumn{4}{c}{[CaII]~7291 $v_{wind}$} & 
\multicolumn{4}{c}{[CaII]~7323 $v_{wind}$} 
}
\decimals

\startdata
AA Tau &                               -164.79 &                                   10.22 &                                -170.44 &                                     8.40 &                                -152.39 &                                    10.20 &                                     .... &                                       .... &                                     .... &                                       .... \\
DL Tau &                               -236.64 &                                    9.57 &                                    .... &                                      .... &                                    .... &                                      .... &                                     .... &                                       .... &                                     .... &                                       .... \\
DQ Tau &                               -140.74 &                                   11.44 &                                    .... &                                      .... &                                    .... &                                      .... &                                     .... &                                       .... &                                     .... &                                       .... \\
FN Tau &                               -125.50 &                                   35.53 &                                    .... &                                      .... &                                    .... &                                      .... &                                     .... &                                       .... &                                     .... &                                       .... \\
Haro 6-13 &                               -194.49 &                                    6.31 &                                    .... &                                      .... &                                    .... &                                      .... &                                     .... &                                       .... &                                     .... &                                       .... \\
HK Tau A &                                -182.88 &                                  .... &                                    .... &                                      .... &                                    .... &                                      .... &                                     .... &                                       .... &                                     .... &                                       .... \\
HK Tau B &                                -56.49 &                                    4.55 &                                    .... &                                      .... &                                    .... &                                      .... &                                     .... &                                       .... &                                     .... &                                       .... \\
HL Tau &                               -140.74 &                                    9.11 &                                -132.43 &                                    31.01 &                                -152.25 &                                    34.56 &                                 -146.17 &                                     24.44 &                                 -107.96 &                                      9.82 \\
IQ Tau &                               -169.29 &                                    7.54 &                                    .... &                                      .... &                                    .... &                                      .... &                                     .... &                                       .... &                                     .... &                                       .... \\
UY Aur &                               -163.31 &                                   10.92 &                                    .... &                                      .... &                                    .... &                                      .... &                                     .... &                                       .... &                                     .... &                                       .... \\
GI Tau &                               -189.32 &                                   17.51 &                                    .... &                                      .... &                                    .... &                                      .... &                                     .... &                                       .... &                                     .... &                                       .... \\  
\enddata
\tablenotetext{}{\textnormal{Note:} \textnormal{The unit of the wind velocity shown here is in $km~s^{-1}$.}
}
\label{tab:OI6300_velocity}
\end{deluxetable*}
\clearpage

\end{document}